\documentclass{aastex}

\newcommand{\be}{\begin{eqnarray}}
\newcommand{\ee}{\end{eqnarray}}

\newcommand\gcm{g~cm$^{-3}$}
\newcommand\simgreater{\,\lower0.7ex\hbox{$\stackrel{>}{\sim}$}\,}
\newcommand\simless{\,\lower0.7ex\hbox{$\stackrel{<}{\sim}$}\,}
\newcommand\msol{$M_\odot$}

\newcommand{\nue}{$\nu_{{\rm e}}$}
\newcommand{\nuebar}{$\bar{\nu}_{{\rm e}}$}
\newcommand{\nux}{$\nu_{{\rm x}}$}

\newcommand{\mgfld}{{\small MGFLD}}
\newcommand{\mgfldtrans}{{\small\sl MGFLD-TRANS}}
\newcommand{\boltzagile}{{\small\sl BOLTZTRAN-AGILE}}

\usepackage{graphicx}

\begin{document}

\title{General Relativistic Effects in the Core Collapse Supernova Mechanism}

\author{S. W. Bruenn\altaffilmark{1} and K. R. De Nisco}
\affil{Department of Physics, Florida Atlantic University, Boca Raton, FL 33431--0991}

\and

\author{A. Mezzacappa\altaffilmark{1} }
\affil{Physics Division, Oak Ridge National
Laboratory, Oak Ridge, TN 37831--6354}

\altaffiltext{1}{Institute for Theoretical Physics, University of California 
at Santa Barbara, Santa Barbara, CA 94913}

\begin{abstract}
We apply our recently developed code for spherically symmetric, fully general relativistic (GR)
Lagrangian hydrodynamics and multigroup flux-limited diffusion (\mgfld) neutrino transport to
examine the effects of GR on the hydrodynamics and transport during collapse, bounce, and the
critical shock reheating phase of core collapse supernovae. GR effects were examined by
performing core collapse simulations from several precollapse models in the Newtonian limit,
in a hybrid limit consisting of GR hydrodynamics and Newtonian transport, and in the fully GR
limit. Comparisons of models computed with GR versus Newtonian hydrodynamics show that collapse
to bounce takes slightly less time in the GR limit, and that the shock propagates slightly
farther out in radius before receding. After a secondary quasistatic rise in the shock radius,
the shock radius declines considerably more rapidly in the GR simulations than in the
corresponding Newtonian simulations. During the shock reheating phase, core collapse computed
with GR hydrodynamics results in a substantially more compact structure from the center out to
the stagnated shock, the shock radius being reduced by a factor of 2 after 300 ms for a 25
\msol\ model and 600 ms for a 15 \msol\ model, times being measured from bounce. The inflow speed
of material behind the shock is also increased by about a factor of 2 throughout most of the
evolution as a consequence of GR hydrodynamics. Regarding neutrino transport, comparisons show
that the luminosity and rms energy of any neutrino flavor during the shock reheating phase
increases when switching from Newtonian to GR hydrodynamics. This arises from the close coupling
of the hydrodynamics and transport and the effect of GR hydrodynamics to produce more compact
core structures, hotter neutrinospheres at smaller radii. On additionally switching from
Newtonian to GR transport, gravitational time dilation and redshift effects decrease the
luminosities and rms energies of all neutrino flavors. This decrease is less in magnitude than
the increase in neutrino luminosities and rms energies that arise when switching from Newtonian
to GR hydrodynamics, with the result that a fully GR simulation gives higher neutrino
luminosities and harder neutrino spectra than a fully Newtonian simulation of the same
precollapse model. We conclude with a discussion of some implications of these results regarding
the development of neutrino-driven convection during the shock reheating phase, and supernova
nucleosynthesis.
\end{abstract}


\section{Introduction}
\label{sec:intro}

Realistic simulations of stellar core collapse, by which we include the core collapse supernova
mechanism, the collapse of stellar cores to black holes, the cooling and deleptonization of
nascent neutron stars, the modeling of neutrino-driven winds, and the prediction of neutrino
signatures, requires the implementation of accurate numerical radiation-hydrodynamics over
extremes of temperature, density, and neutrino luminosities. The wide range of outcomes obtained
by different groups engaged in simulating the core collapse supernova mechanism, for example, is
testimony to the difficulty of numerically modeling this multicomponent and highly nonlinear
phenomenon. 

In this paper, we draw attention to the importance of incorporating general relativity in
simulations of stellar core collapse, with particular focus on the core collapse supernova
mechanism. General relativity must be an essential component in the realistic  modeling of this
mechanism because of the very strong gravitational fields that arise. (Hereafter GR will be used to
denote both the noun ``general relativity'' and the adjective ``general relativistic.'') The core
collapse supernova mechanism is still unknown and involves a complex interplay of hydrodynamics and
neutrino transport, both of which can be significantly modified by GR effects in the crucial period
during which the explosion develops. For example, GR effects can  be expected to substantially modify
the hydrodynamics of the core at high densities. An extreme example of this, of course, is the
possibility that in the GR limit one can have continued collapse and the formation of an event horizon.
The neutrino transport will also be modified by GR, directly through redshift, time dilation, curvature,
and aberration effects, and indirectly through its strong coupling to the GR modified hydrodynamics. 

Other aspects of stellar core collapse will benefit from improved radiation-hydrodynamics codes. 
The detection of neutrinos from supernova 1987A \citep{bionta87,hirata87} opened
up a new window on the Universe, through which we can observe deep inside the core of a massive 
star as it goes through its death throes. There are now a number of large underground particle
detectors, currently on line, such as Super Kamiokande (SuperK)
\citep{totsuka92,nakamura94,nakamura94a} and the Sudbury Neutrino Observatory (SNO)
\citep{ewan92,ewan96,sur94,moorhead97}, or detectors being constructed that will be very
sensitive to the neutrino radiation from a core collapse occurring within the Milky Way galaxy,
and sensitive enough to detect neutrinos from a core collapse occurring anywhere within the
Local Group. The information coded in the neutrino fluxes from a Galactic supernova will be
voluminous, shedding light on the supernova mechanism, and possibly on the fundamental
properties of neutrinos themselves, i.e., whether they have masses and are mixed. Simulations of
the energetics and timescales of the neutrino emission in all flavors, incorporating GR
hydrodynamics and multigroup neutrino transport, will be essential for extracting important
features concerning the physics of stellar core collapse and the fundamental properties of
neutrinos from the neutrino signatures recorded by the new terrestrial detectors.

Previous work at various levels has been done on the problem of coupling the GR equations of
gravity, hydrodynamics, and radiation transport in spherical symmetry. \citet{lindquist66}
obtained the general form of the GR Boltzmann equation and its explicit form and the form of
the equations for the first two energy integrated angular moments in a spherical coordinate system 
in a diagonal gauge (i.e., with a zero shift vector). (This was the coordinate system used, for
example, by \citet{misners65} and by \citet{mayw66} in their derivation of the GR 
hydrodynamics equations.) These equations were also presented by \citet{baronmcv89}, who
additionally presented the explicit form of the equations for the first two monochromatic angular 
moments. \citet{mayle85}, \citet{schinder88}, and \citet{mezzacappam89} derived the GR Boltzmann
equation for nondiagonal gauges, which have desirable singularity-avoidance properties;
\citet{schinder88} additionally derived the monochromatic and energy-integrated neutrino distribution
zeroth and first angular moment equations. 

The first fully GR numerical simulation of stellar core collapse, with the aim of modeling the
supernova mechanism, was performed by \citet{schwartz67}, who coupled GR hydrodynamics in the
diagonal gauge with equilibrium diffusion for the neutrino transport. This was followed by more elaborate
stellar core collapse simulations by \citet{wilson71}, who developed a code coupling GR hydrodynamics
and GR Boltzmann transport, again  in the diagonal gauge, but utilizing much simplified (compared to
more recent simulations)  neutrino physics (e.g., all neutrino scatterings were treated as absorptions).
Other codes were subsequently constructed to simulate stellar core collapse with some implementation of
GR---\citet{bowersw82b} (post-Newtonian hydrodynamics, some order $v/c$ effects in neutrino
transport), \citet{vanriperl81,vanriper82} (GR Hydrodynamics), \citet{bruenn85} (GR Hydrodynamics,
transport of order $v/c$), \citet{mayle85} (post-Newtonian hydrodynamics, gravitational redshifting of
neutrinos), \citet{myrab89} (GR hydrodynamics, gravitational redshifting of neutrinos), 
\citet{coopersteinvb86} (hydrodynamics and neutrino transport to order $v/c$),
\citet{fryerhbc99,fryer99} (GR hydrodynamics), \citet{bruenndm00} (GR hydrodynamics and transport)
\citet{liebendorfermt00} (GR hydrodynamics and transport). 

While the above indicates that a number of codes have been constructed which implement GR at various
levels of approximation (and which implement neutrino transport at various levels of approximation),
direct comparisons between Newtonian and GR simulations are few. An exception to this occurred during
the period from the late 1970's through most of the 1980's. During this period a viable explosion
mechanism, referred to as the ``prompt mechanism'', was thought to be a purely hydrodynamic one wherein
the collapsing core on regaining dynamic stability at nuclear densities overshoots its new equilibrium
position and then rebounds to a larger radius. During rebound, the expanding core acts like a spherical
piston and generates a shock wave which propagates into the overlaying matter. While the generation of a
shock wave on core bounce is a generic feature of core collapse, the prompt mechanism requires that
this shock be strong enough to directly propagate outward through the mantle, reverse the velocity of
the material there, and eject the outer layers of the star explosively, thereby ``promptly'' causing the
supernova explosion. Critical to the prompt mechanism is the initial strength of the bounce shock and
the mass of the outer core through which it must propagate before entering the mantle. This, in turn,
was found to depend sensitively on the equation of state and on the use of GR hydrodynamics
\citep{vanriper78, vanripera78,vanriper79,baronck85,baronbbck87,vanriper88,swestylm94}. Comparisons of
Newtonian versus GR simulations of core infall and bounce show that the stronger
``gravitational attraction'' of GR and its tendency to reduce the difference above nuclear density
between the effective adiabatic index of the matter and the critical adiabatic index causes the GR
simulations to produce higher density bounces, larger rebound amplitudes, and stronger initial
shocks.

The failure of the prompt mechanism to produce explosions when modeled with accurate neutrino
transport (except possibly for a very limited region of core masses and equation of state parameters)
and the inability of spherically symmetric simulations to produce explosions at later times 
caused much of the research on the core collapse supernova mechanism in the early to mid 1990's to
shift to the role of multidimensional effects, and in particular, convection. This focus on convection
was motivated in part by the  potential of convection for increasing the likelihood of explosions, and
in part by a number of observations  of SN 1987A, which indicate that extensive mixing occurred
throughout much of the ejected  material, pointing, {\it by inference}, to fluid instabilities arising
during the explosion itself. These investigations have employed multidimensional simulations with
Newtonian hydrodynamics and transport (\citealp*{herantbc92}; \citealp{herantbhfc94};
\citealp*{millerwm93}; \citealp{jankam95,jankam96}; \citealp*{burrowshf95};\citealp{mezzacappacbbgsu98a,
mezzacappacbbgsu98b}). While this work is extremely important, it has ignored the role of GR. 

Recent developments in core collapse supernovae theory (e.g., the possibility of r-process
nucleosynthesis in neutrino driven winds, further investigations of convection and rotation) have
resulted in a growing number of simulations performed with the incorporation of various levels of GR.
This, together with the restriction of previous comparisons of Newtonian and GR simulations to infall
and bounce, and to the hydrodynamic sector of these simulations, has motivated us to extend these
comparisons to the core evolution well beyond bounce and to the neutrino transport sector of the
simulations as well.

Two fully GR radiation-hydrodynamic codes have recently been developed which implement multigroup
neutrino transport. One code, \boltzagile, developed by \citet{liebendorfer00}, couples GR
hydrodynamics, GR Boltzmann neutrino transport, and the Einstein equations. This code is a marriage of
an adaptive mesh, conservative GR hydrodynamic code \citep{liebendorfer00} with a three-flavor
Boltzmann neutrino transport solver \citep{mezzacappab93b,mezzacappab93c} extended to GR. The other
code, \mgfldtrans, developed by \citep*{bruenndm00}, couples GR multigroup flux-limited diffusion
(\mgfld) to Lagrangian GR hydrodynamics. Both codes give qualitatively similar results for core
collapse simulations, and quantitative comparisons are in progress. 

Our purpose in this paper is to present results of simulations performed with \mgfldtrans\ of the
collapse, bounce, and subsequent ($\sim$ 1 second) evolution of the cores of massive stars. These
simulations will be compared with Newtonian simulations, performed with the Newtonian limit of this
same code, in order to assess the effects of general relativity for the evolution of the stellar core,
particularly during the critical phase in which the shock stalls and the success or failure
of the supernova explosion is thereafter by a competition between the neutrino heating behind the shock
and the accretion ram. The code itself will be described in detail in another paper \citep{bruenndm00},
but we will give a further very brief description below.

\mgfldtrans\ solves the GR hydrodynamics equations, the GR \mgfld\ transport equations, and the
Einstein equations. A 3 + 1 formalism using a diagonal gauge is used to advance the equations in time.
The metric used is 

\begin{equation}
ds^{2} = a^{2}(t,m) c^{2} dt^{2} - b^{2}(t,m) dm^{2} - r^{2}(t,m)( d\theta^{2} + \sin^{2} \theta 
d\phi^{2} )
\label{eq:p1}
\end{equation}

\noindent where $m$ is the total rest mass enclosed by a shell of radius r, and $a$ and $b$ are
metric functions. Operator splitting is used to couple the hydrodynamics and transport codes and,
very roughly speaking, is implemented by advancing first the matter configuration and metric
functions through a time step $t \rightarrow t + \Delta t$ using the GR hydrodynamics (using the
neutrino fields at time $t$), and then advancing the neutrino distribution through the same time
step with the GR \mgfld\ code.  The code is capable of performing 1D fully GR simulations of all
phases of stellar core collapse and core collapse supernovae. Because of the sensitivity of the supernova
mechanism to the neutrino spectra, we believe it is essential that numerical simulations be performed
with {\it multigroup} neutrino transport. In this way, the neutrino spectrum is computed {\it ab initio},
rather than assumed, and the critical energy-dependent neutrino interactions can therefore be
calculated accurately. 

The plan of this paper is as follows: In Section \ref{sec:paradigm} we give a brief account of
the current core collapse supernova paradigm in order to set the stage for this work. We then outline
in Section \ref{sec:method} our numerical methods and the simulations we perform. Because matter
hydrodynamics and neutrino transport are closely coupled during stellar core collapse, we compare
Newtonian and GR hydrodynamics in Section \ref{sec:hydro}. This provides a basis for distinguishing
between the modifications of neutrino transport that arise ``directly'' from the GR terms in the
transport equations and those that arise from the effects of GR on the hydrodynamics. Neutrino
transport is our focus in Section \ref{sec:trans}, where we investigate the effects of GR by comparing
dynamical simulations that use various combinations of Newtonian and GR hydrodynamics and transport. We
state our conclusions in Section \ref{sec:cncl}.

\section{Supernova Paradigm}
\label{sec:paradigm}

To set the stage for this work, we very briefly review the current core collapse supernova 
paradigm, which is referred to as the ``shock reheating mechanism'' or ``delayed mechanism''
\citep{wilson85,bethew85}. It is based on the original \citet{colgatew66} idea that a core
collapse supernova explosion is driven by neutrino energy deposition, but the paradigm has been
much modified and refined over the intervening years. All investigators agree with the paradigm's
account of the initial phases of the mechanism, which starts with the destabilization and collapse
of the core of a massive star. When the density of the core exceeds nuclear matter density, the
homologously collapsing inner core rebounds and drives a shock into the outer core. This
``bounce'' shock, which must ultimately explode the star, weakens and stalls between 100 and 200
km. This temporary ``demise'' of the shock is brought about by the reduction in the postshock
pressure due to both the dissociation of nuclei in the material passing through by the shock and
the intense outward radiation of neutrinos from the postshock region. The stalled shock becomes an
accretion shock, separating the supersonically infalling material at larger radii from the
material at smaller radii, which is subsonically settling onto the surface of the proto-neutron
star. Within 10's of milliseconds, the structure of the core becomes quasi--steady-state. Infalling
matter encounters the shock and is shock dissociated into free nucleons. As these nucleons
continue to flow subsonically inward, they are heated by absorbing a small fraction of the
neutrinos that are being radiated from the hot contracting core. As this accreting matter continues
to flow inward, neutrino and compressional heating increase its temperature until the cooling
rate, which goes as the sixth power of the temperature, exceeds the heating rate. The inflowing
matter thereafter cools and ultimately accretes onto the core. An important radius that we will
have a number of occasions to refer to is the ``gain radius,'' at which the heating and cooling
rates balance. Above the gain radius, heating by the absorption of neutrinos dominates; below the
gain radius, cooling by the emission of neutrinos dominates. (Gain radii can be defined
analogously for each neutrino flavor, e.g., the \nue\ gain radius is the radius at which \nue\
heating and \nue\ cooling vanish.)

It is here that the shock reheating mechanism becomes murky, as different groups fail to agree on
when, how, or even if the shock is revived. The problem is that simulating this epoch realistically
presents many difficulties. Neutrino transport plays a key role, for example, but until now the
state of the art in neutrino transport algorthms has been multigroup flux-limited diffusion. In
this scheme, the neutrino spectrum is computed as part of the transport solution rather than being
assumed, which is very important, but the diffusion and free streaming limits are bridged by
interpolation. Unfortunately, this interpolation encompasses the critical neutrino heating region
between the gain radius and the shock. Furthermore, a generic problem with flux-limited diffusion
is that it underestimates the isotropy of the neutrino distribution when the transition from large
to small neutrino opacities is abrupt, as it is near the neutrinospheres shortly after the
formation of the proto-neutron star. [Fortunately, ``exact'' Boltzmann transport simulations of
neutrino transport are becoming available
\citep{mezzacappab93a,mezzacappab93b,mezzacappab93c,messermbg98,mezzacappalmhtb00}.] Another
difficulty is that the neutrino heating between the gain radius and the shock sets up an unstable
entropy gradient that drives convection. This convection, referred to as neutrino-driven
convection, may help in reviving the shock, but can only be modeled realistically by
multidimensional (preferably 3-dimensional) simulations. It is also likely that fluid instabilities
also arise below the neutrinospheres, and these may drive fluid motions which, by advecting
entropy and leptons to or from the neutrinospheres, may modify the neutrino transport. It is not
clear yet whether these fluid instabilities above and below the neutrinopsheres play an essential
or a peripheral role in the explosion mechanism. In any case, the radius of the shock is determined
by the competition between the thermal and convective pressure generated by neutrino heating, on
the one hand, and by the ram pressure of the infalling material on the other. The generation of a
successful explosion requires that this radius becomes unstable
\citep{burrowsg93} so that quasi-hydrostatic readjustment becomes impossible, i.e., that the
neutrino heating be sufficiently rapid in the region between the gain radius and the shock for the
thermal and convective pressure to overcome the accretion ram, causing the shock to accelerate
outwards. Unfortunately, explosions seem to be marginal at best in many of the detailed numerical
simulations, and perhaps in Nature as well; therefore, they depend sensitively on the
implementation of the physics and numerics: e.g., the equation of state, neutrino opacities,
hydrodynamics, neutrino transport, and general relativity. In particular, realistic simulations of
core collapse supernovae must include the effects of GR.

\section{Method}
\label{sec:method}

To assess the GR effects on the evolution of the core in the context of the above core
collapse supernova paradigm, we have performed two sets of simulations, one set initiated
from a 15 \msol\ precollapse model, S15s7b, the other set initiated from a 25
\msol\ precollapse  model, S25s7b. These precollapse models were provided by
\citet{woosley95}, and are described in \citet{woosleyw95}. The precollapse models S15s7b and
S25s7b were chosen to represent small-iron-core and large-iron-core precollapse  models,
respectively. When core collapse begins, the iron core mass of model S15s7b is 1.278 \msol,
while that of model S25s7b is 1.770 \msol. 

The set of simulations initiated from model S15s7b consists of a simulation with Newtonian
hydrodynamics and Newtonian transport, referred to as simulation S15s7b\_nt, a hybrid
simulation with GR hydrodynamics and Newtonian transport, referred to as simulation
S15s7b\_hyb, and a fully GR simulation, referred to as simulation S15s7b\_gr. Likewise, the
set of simulations initiated from model S25s7b consists of the same combinations of Newtonian
and GR hydrodynamics and transport, and are referred to as S25s7b\_nt, S25s7b\_hyb,
and S25s7b\_gr. In each simulation, the model was evolved through core collapse, bounce, and
through an 0.8 to 1.0 second interval after bounce. (The simulations beginning with S25s7b
with GR hydrodynamics (viz., simulations S25s7b\_hyb and S25s7b\_gr) resulted in the
formation of an event horizon $\sim$ 0.6 seconds after bounce, and had to be terminated.) For
the simulations carried out with Newtonian transport (viz., simulations S15s7b\_nt,
S15s7b\_hyb, S25s7b\_nt, and S25s7b\_hyb), the transport was computed using the
Newtonian limit of our fully GR \mgfld\ code,
\mgfldtrans, described briefly in Section \ref{sec:intro}. In this limit, the code is
essentially as described in \citet{bruenn85} and \citet{bruennh91}. We will refer to this
transport as ``Newtonian \mgfld'', even though, technically, the transport equations in this
limit are derived from the $O(\frac{v}{c})$ Boltzmann equation. For the simulations carried
out with GR transport (viz., simulations S15s7b\_gr and S25s7b\_gr), the neutrino transport
was computed with \mgfldtrans\ in its fully GR mode.

With the above described simulations, we will compare Newtonian and GR hydrodynamics as they
affect such quantities as the locations of the shock, the \nue\ and \nuebar\ gain radii, and
the infall speed of the matter between the shock and the gain radii (and therefore the time
available for a comoving fluid element to be heated by neutrinos).  We will also compare
Newtonian and GR transport as they affect such quantities as the luminosities and rms
energies of each neutrino flavor.  Comparisons between the Newtonian and hybrid simulations
will highlight the effects on the neutrino transport of the differences between Newtonian and
GR hydrodynamics. On the other hand, the hydrodynamics in the hybrid and the fully GR
simulations is the same, and comparisons between these simulations will highlight the
differences between Newtonian and GR neutrino transport. 

In all simulations, the Lattimer-Swesty equation of state (Lattimer \& Swesty 1991) was used
when the following three conditions were satisfied locally: (1) $n_{{\rm B}} > 10^{-8}$
fm$^{-3}$  ($\rho > 1.67 \times 10^{7}$ \gcm), where $n_{{\rm B}}$ is the number density of
nucleons (free  and bound) per cubic Fermi, (2) $T > 0.05$ MeV, and (3) the matter was
assumed to be in nuclear statistical equilibrium (NSE). The  Cooperstein-BCK equation of state
(\citealp{cooperstein85a}; \citealp*{baronck85}) was used when the second and third condition
was satisfied, but not the first. For nuclei not in NSE, which comprised the silicon layer,
oxygen layer, and other exterior layers until they encountered the shock, the nuclei were
treated as an ideal gas (with excited states), and a nine-species nuclear reaction network
was used to follow the nuclear transmutations. A zone not in NSE were flashed to NSE if
its temperature reached 0.44 MeV \citep*{thielemannnh96}. Three-flavor GR
multigroup flux-limited diffusion, as described in \citep{bruenndm00}, was used for the
neutrino transport, with twenty energy zones spanning in geometric progression the neutrino
energy range from 4 MeV to 400 MeV. The neutrino microphysics described in
\citet{bruenn85} and \citet{bruennh91} was used, as well as the ion screening corrections
given by \citet{horowitz96}, described in detail in \citet{bruennm97}. 

\section{General Relativistic versus Newtonian Hydrodynamics}
\label{sec:hydro}

The structure of the core during the shock reheating epoch consists of a quasi-stationary 
accretion shock separating the supersonic infall of mantle material outside the shock from the 
subsonic inward flow of material between the shock and the surface of the proto-neutron  star.
This structure is similar to that of a neutron star undergoing hypercritical accretion (i.e.,
accretion far in excess of the photon Eddington limit), which can occur after the supernova
explosion if some of the ejecta falls back onto the neutron star, or if a neutron star spirals
into the envelope of a binary companion. Hypercritical accretion has been investigated by
\citet{colgate71}, \citet*{zeldovichin72}, \citet{blondin86}, \citet{chevalier89},
\citet{chevalier95}, and \citet{houckc91}. As emphasized by \citet{houckc91} and
\citet{chevalier95}, the steady-state position of the shock is determined by the condition that
the neutrino radiation, which occurs in a thin layer around the neutron star surface, release the
gravitational binding energy gained by the infalling material. The neutrino radiation occurs in a
thin layer because of the sensitivity of the neutrino emissivity to the temperature. The
dependence is $T^{9}$ for the pair process, applicable for the wide range of expected accretion
rates ($10^{-8}$ -- $10^{4}$ \msol/yr) after the supernova, and
$T^{6}$ for electron and positron capture, applicable for the much higher accretion rates (and
densities) during the shock reheating epoch. 

\citet{houckc91} studied hypercritical accretion onto a neutron star with GR hydrodynamics and
found that the shock radius was reduced by a factor of about two from the value given by Newtonian
hydrodynamics. Their explanation for this GR reduction in shock radius applies to a similar
reduction in shock radius found in our simulations of the shock reheating epoch. The explanation
begins with the fact that the pressure at the neutron star surface is fixed by the temperature
required for neutrinos to radiate away the gravitational binding energy of the infalling 
material. At a given accretion rate, more binding energy must be radiated away with GR than with
Newtonian hydrodynamics because of the more compact core structure and deeper gravitational
potential well that results with GR. On the other hand, the high temperature sensitivity of the
neutrino emission rates leads to temperatures in the two calculations, and therefore pressures,
that do not differ greatly at the emission surface despite the difference in the emission rates.
Above the emission surface, however, the stronger effective gravitational field strength in the GR
calculation results in a larger pressure gradient. The approximately equal pressures at the
neutron star surfaces in the two calculations together with a larger pressure gradient in the GR
calculation results, in the latter calculation, in a more compact structure between the neutron
star surface and the shock. 

Figures \ref{fig1} and \ref{fig2} compare the shock radius and the \nue\ and \nuebar\ gain radii as
a function  of time for the postbounce evolution of models S15s7b and S25s7b, respectively, as
given by Newtonian and GR hydrodynamics for the same (Newtonian) transport, i.e., simulations
S15s7b\_nt, S15s7b\_hyb, S25s7b\_nt, and S25s7b\_hyb. We mention again that the \nue\ gain radius
for the inflowing material is the radius at which cooling by \nue\ emission is equal to heating by
\nue\ absorption.  Above the gain radius, \nue\ heating dominates; below the gain radius, \nue\
cooling  dominates. The \nuebar\ gain radius is defined analogously, and, as in the case of the
\nue's, \nuebar\ heating and cooling dominate above and below the \nuebar\ gain radius,
respectively.  The energy transfer between neutrinos and matter behind the shock during the
postshock evolution  considered here is mediated primarily by the charged current reactions

\begin{equation}
\nu_{\rm e} + {\rm n} \rightleftharpoons {\rm p} +{\rm e^{-}}, \hspace{0.5in}
\bar{\nu}_{\rm e} + {\rm p} \rightleftharpoons {\rm n} + {\rm e^{+}}.
\label{eq:p2}
\end{equation}

\noindent Although $\mu$ and $\tau$ neutrinos are also emitted by the proto-neutron star at this 
time, they couple very weakly with the material above the proto-neutron star and, therefore, play 
a negligible role in matter heating. We omit their gain radii in Figures \ref{fig1} and
\ref{fig2}. 

It is seen from the figures that, as in the case of hypercritical accretion onto a fully  formed
neutron star, the structure above the proto-neutron star during the shock reheating epoch  is
considerably more compact in the GR simulations than in the Newtonian simulations. The shock
radius is reduced by a factor of about 2 after $t_{pb} \sim 0.3$~s for simulation S25s7b\_hyb
versus simulation S25s7b\_nt, and after $t_{pb} = 0.6$~s for simulation S15s7b\_hyb versus
simulation S15s7b\_nt. Throughout most of the postbounce evolution, the distance between the gain
radius and the shock is also reduced by a factor of about 2 for the simulations with GR
hydrodynamics. This constriction of the heating region exhibited by the simulations with GR
hydrodynamics will result in a reduction in the time spent in this region by inwardly moving fluid
elements, and therefore, for given neutrino luminosities and rms energies, in a reduction in the
net heat acquired by these fluid elements. 

In addition to the width of the heating region, the material inflow speed through the heating 
region affects the net heat acquired by the material, and is modified by the use of GR
hydrodynamics. We note first that the material inflow speed through the heating region below the
shock is a very shallow function of the radius. This behavior can be understood from the fact that
the flow below the shock is subsonic, approximately adiabatic until intense cooling sets in  near
the neutrinospheres, and has an approximately constant $\gamma$, where $\gamma$ is the adiabatic
index. Under these conditions \citep{chevalier89},
 
\begin{equation} v \propto r^{( 3 - 2\gamma )/( \gamma - 1 )} ,
\label{eq:p4}
\end{equation}

\noindent which follows by integrating the equation of hydrostatic equilibrium and the continuity
equation through this region. Now the material below the shock is a mixture of radiation, leptons,
and free baryons, with a value of gamma between 1.4 and 1.5. Equation (\ref{eq:p4}) shows that $v$
is indeed a rather shallow function of $r$ for this range of gamma. We can therefore characterize
the speed of the material through the heating region by its immediate postshock velocity. Figure
\ref{fig3} compares the postshock velocities as a function  of time for simulations S15s7b\_nt,
S15s7b\_hyb, S25s7b\_nt, and S25s7b\_hyb. It is seen that the postshock velocities given by the
simulations with GR hydrodynamics, i.e., simulations S15s7b\_hyb and S25s7b\_hyb, are substantially
larger in magnitude than those given by the simulations with Newtonian hydrodynamics, i.e.,
simulations S15s7b\_nt and S25s7b\_nt, the difference being $\sim$ 2 for each model after roughly
0.2 s. The larger postshock velocities in the GR simulations follows from the fact that (a) the
immediate preshock velocities in the GR limit are about a factor of 2 larger than those given in
Newtonian limit because of the smaller shock radii and the stronger effective gravity in the GR
case, and (b) the fact that the ratios of the velocities across the shock are approximately the
same for both the Newtonian and GR simulations. The differences between GR and Newtonian
hydrodynamics, in particular, the much reduced time that a given fluid element spends between the
shock and the gain radius in the GR simulations, may have implications for the development of
neutrino-driven convection.

\section{General Relativistic versus Newtonian Hydrodynamics and Transport}
\label{sec:trans}

In this section, we will compare complete dynamic simulations of core collapse with Newtonian
hydrodynamics and transport, GR hydrodynamics and Newtonian transport, and GR hydrodynamics and
transport. 

\subsection{Infall and Bounce}
\label{sec:infall_bounce}

We will begin our comparisons with the infall, bounce, and immediate postbounce epochs as computed
with Newtonian hydrodynamics and transport, viz., simulations S15s7b\_nt and S25s7b\_nt, and GR
hydrodynamics and transport, viz., simulations S15s7b\_gr and S25s7b\_gr. Figures \ref{fig4} --
\ref{fig7} show radius versus time trajectories of Lagrangian surfaces enclosing selected rest
masses. Figures \ref{fig4} and \ref{fig6} show this for the GR simulations, Figures \ref{fig5} and
\ref{fig7} show this for the corresponding Newtonian simulations. For both precollapse models,
collapse to bounce takes slightly less time in the GR simulations --- 0.1829 s for S15s7b\_gr
versus 0.2013 s for S15s7b\_nt, and 0.2723 s for S25s7b\_gr versus 0.2895 s for S25s7b\_nt. Bounce
occurs at a slightly higher density in the GR simulations --- $4.23 \times 10^{14}$ g/cm$^{3}$ for
S15s7b\_gr versus $3.20 \times 10^{14}$ g/cm$^{3}$ for S15s7b\_nt, and $4.15 \times 10^{14}$ 
g/cm$^{3}$ for S25s7b\_gr versus $3.34 \times 10^{14}$ g/cm$^{3}$ for S25s7b\_nt. For both models,
the shock propagates slightly farther out in radius in the GR simulations before receding --- 172
km for S15s7b\_gr versus 155 km for S15s7b\_nt, and 231 km for S25s7b\_gr versus 226 km for
S25s7b\_nt. After the secondary quasistatic rise in the shock radius, the shock radius declines
considerably more rapidly in the GR simulations than in the corresponding Newtonian simulations. 

Figure \ref{fig8} shows for simulations S15s7b\_gr and S15s7b\_nt the luminosity as a function of
time for \nue's, \nuebar's, and \nux's. Here \nux\ refers to either a $\nu_{\mu}$,
$\bar{\nu}_{\mu}$,  $\nu_{\tau}$, or $\bar{\nu}_{\tau}$. (These have very similar interactions with
matter at the energies of importance in core collapse supernovae, and are treated identically in
the transport code.) Figure \ref{fig9} is similar, plotting the \nue, \nuebar, and \nux\
luminosities as a function of time for simulations S25s7b\_gr and S25s7b\_nt. The spike in the
\nue\ luminosity is produced when the shock propagates out through the \nue-sphere, which at this
time is located at a radius of roughly 100 km for all models. The source of the \nue\ luminosity
spike is (a) the \nue's produced by the rapid electron capture on the free protons released by the
shock-dissociated nuclei, and (b) their production near the \nue-sphere which
 results in their rapid escape from the core. The delay in the \nue-burst for S15s7b\_nt relative
to S15s7b\_gr, and in S25s7b\_nt relative to S25s7b\_gr, is a consequence of the corresponding
delay in the bounce and shock generation in the Newtonian relative to the GR simulations,
described above. In the case of simulations S25s7b\_gr and S25s7b\_nt (Figure \ref{fig9}), there
is a small secondary spike in the \nue\ and \nuebar\ luminosities that occurs $\sim$ 15 ms after
the primary \nue\ spike. These spikes are caused by the recollapse of material that occurs $\sim$
15 ms following bounce, as evident in Figures \ref{fig6} and \ref{fig7}. The recollapse of material
draws in and heats the \nue\ and \nuebar\ neutrinospheres, thus hardening the \nue\ and \nuebar\
spectra (Figure \ref{fig11}) and increasing their luminosities. 

Figure \ref{fig10} shows for simulations S15s7b\_gr and S15s7b\_nt the rms energy as a function of
time for the \nue's, \nuebar's, and \nux's. Figure \ref{fig11} is similar, plotting the \nue,
\nuebar, and \nux\ rms energies as a function of time for simulations S25s7b\_gr and S25s7b\_nt.
These figures show that in all the simulations there is a trifurcation of the rms energies of the
different neutrino flavors which begins immediately after bounce---the \nue's having the smallest
rms energy at any given time, and the \nux's having the largest. This well known effect is caused
by the flavor dependence of the coupling strength of neutrinos with matter, which cause the
different neutrino flavors to thermally decouple from the matter at different radii. The \nue's and
\nuebar's thermally couple with matter primarily by the charged current processes (\ref{eq:p2}).
For a given neutrino energy, the large neutron to proton ratio in the core endows the matter with
higher \nue\ opacity  than \nuebar\ opacity, and the \nuebar's therefore tend to thermally
decouple from the matter deeper in the core where it is hotter. The \nux's lack any charged current
opacity contribution, and they thermally decouple from the matter deepest in the core and thus have
the highest rms energies. 

A feature in these plots that appears in all the simulations is the prominent spike in the rms
energy of the \nux's at bounce. The luminosities of the \nux's at this time are still small, so
the overall number of \nux's affected is small. The spike in the rms energy has its origin in the
compression of the \nux's by the shock immediately before it propagates through the \nux-sphere,
and is numerical artifact of the pseudoviscous terms in the numerical hydrodynamics which enable
the hydrodynamics code to track shock waves automatically, but which spread the shock
compression-front over several radial zones. This numerical artifact arises from the fact that,
unlike the \nue's and \nuebar's, the \nux-opacity is dominated by isoenergetic scattering on
nucleons and nuclei. Much weaker are the processes that thermally equilibrate the \nux's with
matter, i.e., nucleon bremsstrahlung (not included in these simulations), neutrino-electron
scattering, and \nux-$\bar{\nu}_{x}$ pair annihilation. Therefore, the \nux-thermalization-sphere,
defined as having the radius at which the \nux's are last thermally equilibrated with the matter,
lies considerably deeper in the core than the \nux-sphere, defined as having the radius at which
the \nux's decouple from the matter completely. During the brief time that the supernova shock
generated from core bounce propagates outward through the region between the
\nux-thermalization-sphere and the \nux-sphere, the artificially large width of the
pseudoviscous-spread shock compression-front exceeds the \nux-scattering mean-free-paths, and the
\nux's are therefore compressed along with the matter and boosted up in energy. The \nux-spectrum,
which normally reflects the conditions at the \nux-thermalization-sphere, is thus momentarily and
unphysically hardened by this shock-compression.

\subsection{Shock Reheating}

Figure \ref{fig12} shows a comparison of the \nue\ luminosity as a function of time during the
shock reheating epoch for simulations S15s7b\_nt, S15s7b\_hyb, and S15s7b\_gr. Figures \ref{fig13}
and \ref{fig14} are similar to Figure \ref{fig12} and show, respectively, a comparison of the
\nuebar\ luminosity and the \nux\ luminosity for these simulations. Figures \ref{fig15},
\ref{fig16}, and \ref{fig17} show, respectively, a comparison of the \nue, \nuebar, and \nux\
luminosity as a function of time during the shock reheating epoch for the simulations S25s7b\_nt,
S25s7b\_hyb, and S25s7b\_gr. Figures \ref{fig15} and \ref{fig16} show that the \nue\ and \nuebar\
luminosities for all simulations of model S25s7b exhibit an abrupt decline beginning at about 0.42
s after bounce. This occurs when the interface between the silicon and oxygen shells passes through
the neutrinospheres. There is a considerable jump in entropy and a corresponding drop in density
at this interface, and the passage of this interface through the neutrinospheres reduces the mass
accretion rate there by almost a factor of three. This substantially reduces the \nue\ and
\nuebar\ accretion luminosities, and accounts for the above mentioned drop in their overall
luminosities. The \nux\ luminosity arises deeper in the core, has very little contribution from
the accreting matter, and is therefore unaffected by the drop in the mass accretion rate. 

Figures \ref{fig18}, \ref{fig19}, and \ref{fig20} show, respectively, a comparison of the \nue,
\nuebar, and \nux\ rms energy as a function of time from bounce for the simulations S15s7b\_nt,
S15s7b\_hyb, and S15s7b\_gr. Figures \ref{fig21}, \ref{fig22}, and \ref{fig23} do the same for the
simulations S25s7b\_nt, S25s7b\_hyb, and S25s7b\_gr. The neutrino luminosities and the rms
energies exhibit a trend that is common to all neutrino flavors for both models S15s7b and S25s7b.
During the shock reheating phase, the luminosity and rms energy of any neutrino flavor increase
when switching from Newtonian to GR hydrodynamics, and decrease when switching from Newtonian to
GR transport. The increase in the neutrino luminosity and rms energy in switching from Newtonian
to GR hydrodynamics can be understood by recalling that GR hydrodynamics produces a more compact
core structure, as discussed in Section \ref{sec:hydro}. This results in hotter neutrinospheres at
smaller radii. For example, in simulation S15s7b\_nt, at 0.4 s after bounce, the \nue-sphere has a
radius and temperature of 37.7 km and 3.89 MeV, respectively. The corresponding values for radius
and temperature in simulation S15s7b\_hyb are 29.0 km and 4.80 MeV, respectively. At the same
postbounce time, the \nuebar-sphere has a radius and temperature of 35.5 km and 4.15 MeV,
respectively, in simulation S15s7b\_nt, compared to a radius and temperature of 27.0 km and 4.92
MeV, respectively, in simulation S15s7b\_hyb. Moreover, at this time, the
\nux-thermalization-sphere has a radius and temperature of 25.8 km and 6.53 MeV, respectively, in
simulation S15s7b\_nt compared to a radius and temperature of 24.28 km and 7.66 MeV, respectively,
in S15s7b\_hyb.

The decrease in the luminosity and rms energy of all neutrino flavors when switching from Newtonian
to GR transport is primarily a consequence of (a) gravitational redshift, as the neutrinos
propagate out to large distances, and (b) the difference between the rate of proper time at the
respective neutrinospheres and at large radii. The gravitational redshift occurs as a neutrino
propagates from its neutrinosphere to a large radius and affects both the neutrino luminosity and
rms energy. The difference between the rates of proper time causes an additional reduction in the
neutrino luminosity (by the same factor as the gravitational redshift), as it causes an observer at
a large radius to see a reduced rate of photon emission from the neutrinosphere. 

The radii and temperatures of the various $\nu$-spheres are similar for simulations S15s7b\_hyb and
S15s7b\_gr at corresponding times after bounce. The same is true for simulations S25s7b\_hyb and
S25s7b\_gr. They differ from the radii and temperatures given by the corresponding Newtonian
simulations, S15s7b\_nt and S25s7b\_nt, mainly by the tendency, discussed above, of the GR
hydrodynamics to produce a more compact and hotter core structure. Therefore, the main effect on
the radii and temperatures of the various $\nu$-spheres occurs when switching from Newtonian to GR
hydrodynamics. However, there are some other rather small effects on the properties of the
$\nu$-spheres that occur when switching from the hybrid to the fully GR simulations. For example,
as the neutrinos gradually decouple from the matter, they propagate outward with increasingly long
mean free paths. In the simulations with GR transport, they consequently suffer some gravitational
redshifting between successive interactions with the matter before they completely decouple. This
slightly lowers their rms energies, reduces their interaction cross sections, and causes them to
decouple from the matter at a slightly smaller radius. The net effect is a slight reduction (about
0.2 km) in the radii of the neutrinospheres. These small effects will be discussed in
\citet{bruenndm00}, along with a detailed description of the GR transport code.

\section{Discussion and Conclusions}
\label{sec:cncl}

We have developed a general relativistic \mgfld\ neutrino transport code, coupled it to Lagrangian
GR hydrodynamics, and used it to simulate core collapse supernovae. Beginning with several
precollapse models, we have performed purely Newtonian simulations, hybrid simulations using general
relativistic hydrodynamics and Newtonian \mgfld\ neutrino transport, and fully general relativistic
simulations. We have shown that the effect of general relativistic hydrodynamics, versus Newtonian
hydrodynamics, is to make substantially more compact the  structure of the core up to the stagnated
shock: the shock radius is reduced by a factor of 2 for postbounce times exceeding 300 ms for model
S25s7b and 600 ms for model S15s7b. Moreover, the inflow speed of the material behind the shock is
also increased by about a factor of 2 throughout most of the postbounce evolution in these models. 

We have also compared the \nue, \nuebar, and $\nu_{x}$ luminosities and rms energies for the same
three sets of simulations. We find that switching from Newtonian to GR {\em hydrodynamics} increases
the luminosity and rms energy of all neutrino flavors during the shock reheating epoch. This arises
because of the more compact core structures that develop with GR hydrodynamics, which yields
$\nu$-spheres at smaller radii and higher temperatures. The higher $\nu$-sphere temperatures
increase the luminosities and rms energies of all three neutrino flavors. The smaller radii reduce
their luminosities, but not enough to offset the luminosity increase due to the higher temperatures.
Switching from Newtonian to GR {\em transport} reduces the luminosities and rms energies of all three
neutrino flavors during the shock reheating epoch because of the gravitational redshift of the
neutrinos as they propagate out to large radii and because of the difference between the rate of
proper time at the various $\nu$-spheres compared with the rate of proper time at larger radii.
With one exception (the rms \nux\ energies for model S15s7b), the reduction in neutrino luminosities
and rms energies when switching from Newtonian to GR {\em transport} does not fully compensate for
the increase in these quantities when switching from Newtonian to GR {\em hydrodynamics}. Therefore,
the net effect in switching from a fully Newtonian to a fully GR simulation in most cases is an
increase in both the luminosities and rms energies of neutrinos of all flavors during the shock
reheating epoch.

The results described in this paper show that GR effects make substantial changes in both the
structure of the postcollapse core and the neutrino emission (luminosities and spectra) at this
epoch. Therefore, it is important that simulations focused on the supernova mechanism, observables
such as the neutrino signatures in underground neutrino observatories, and the nucleosynthesis
occurring in the neutrino-driven wind immediately following a supernova explosion, be carried out in
the fully GR limit.

\acknowledgments

SWB and KRD are supported at Florida Atlantic University in part by NSF grant AST--9618423
and NASA grant NAG5-3903. AM is supported at the Oak  Ridge National Laboratory,
managed by UT-Battelle, LLC, for the U.S. Department of Energy under contract
DE-AC05-00OR22725. Some of the simulations  presented here were performed on the 
Multidisiplinary Research Computing Facility (MDRCF) at the College of Science at 
Florida Atlantic University, funded in part by FAU and the NSF.

\clearpage

\figcaption[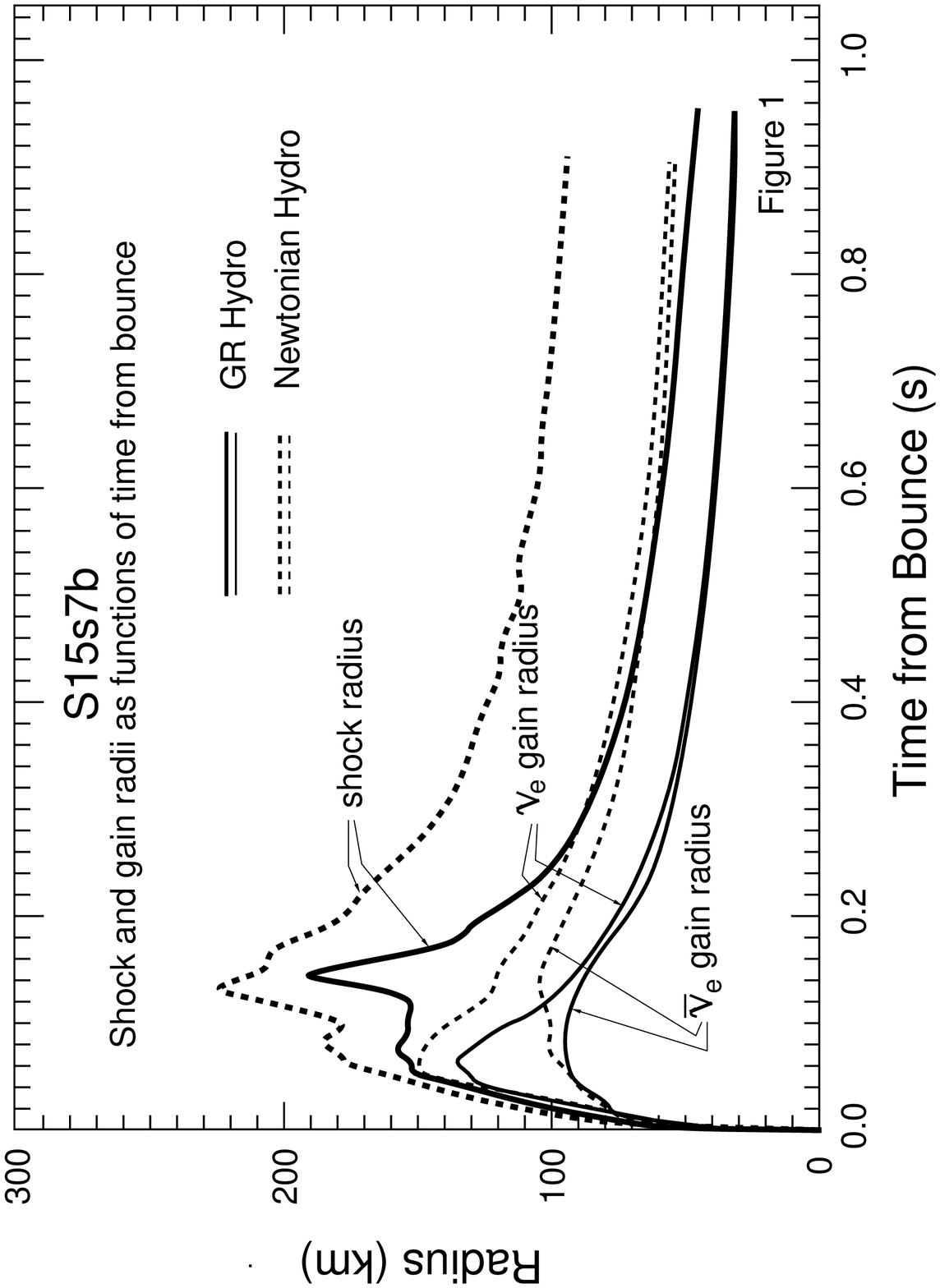] {Shock and gain radii as functions of time from bounce for simulations
S15s7b\_nt and S15s7b\_hyb. The dashed lines refer to the simulation S15s7b\_nt (Newtonian
hydrodynamics), the solid lines refer to the simulation S15s7b\_hyb (GR hydrodynamics). Both
simulations utilize Newtonian transport.
\label{fig1}} 

\figcaption[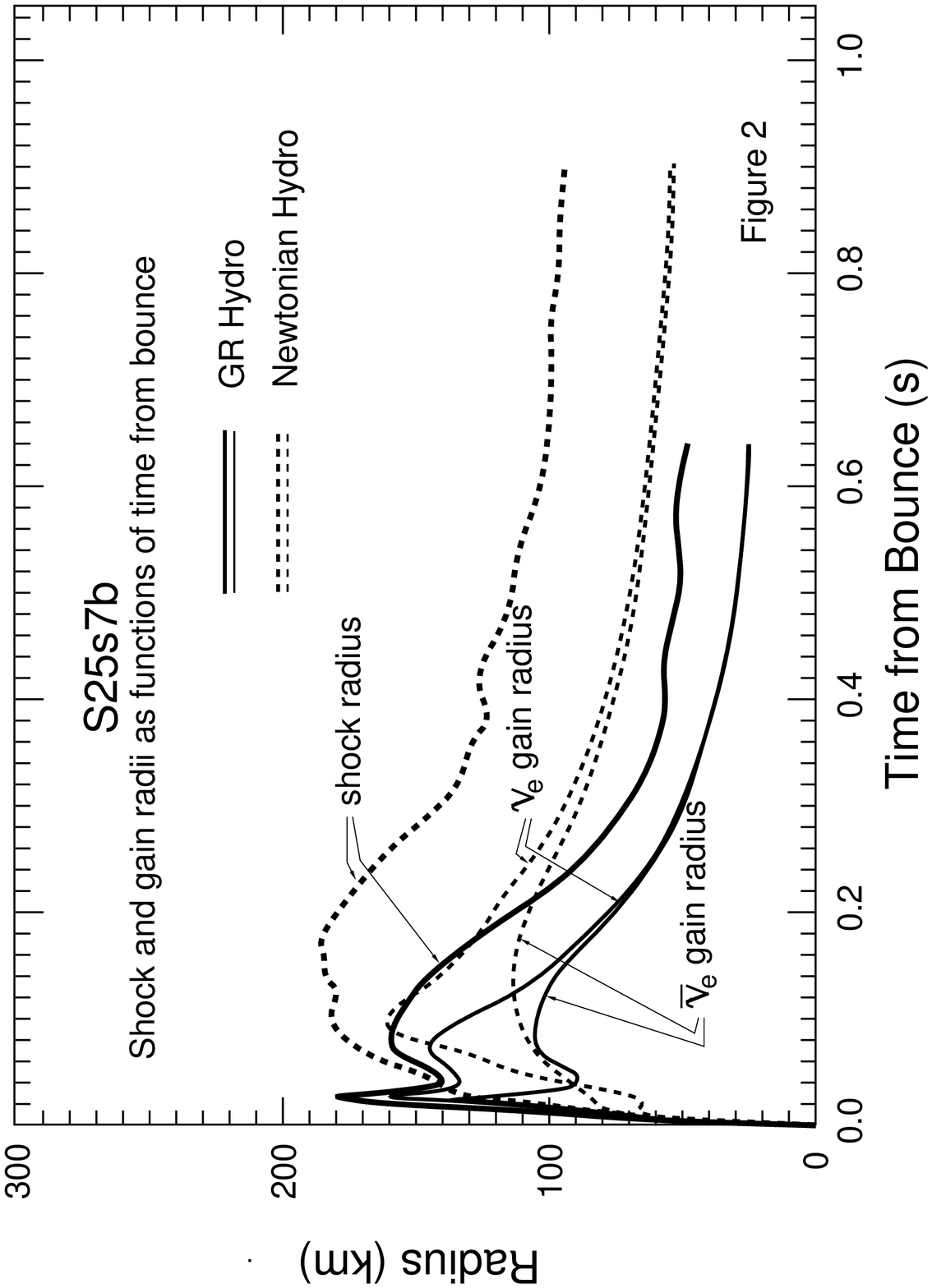] {Same quantities shown in Figure 1, but for simulations S25s7b\_nt and
S25s7b\_hyb..
\label{fig2}} 

\figcaption[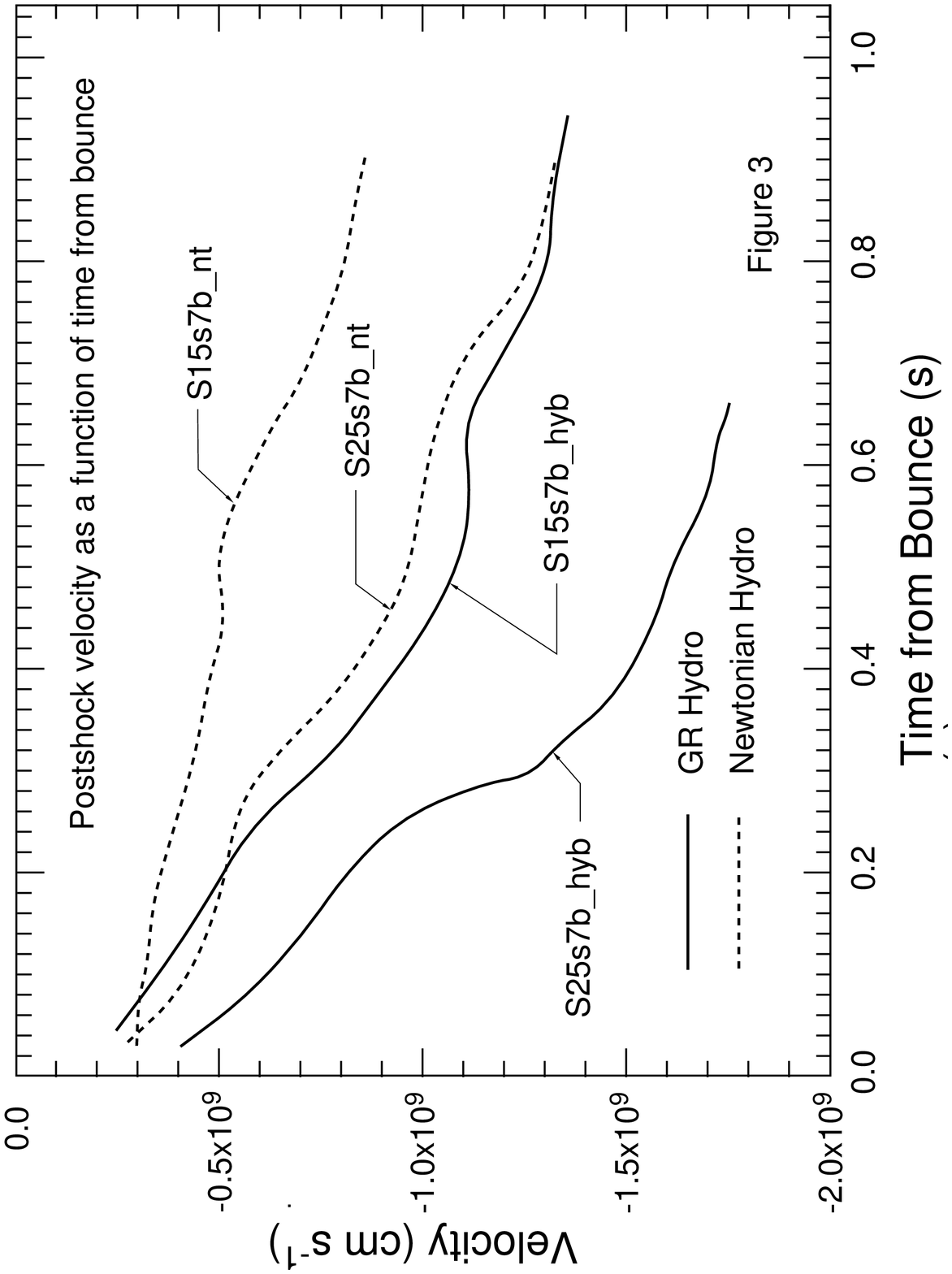] {Postshock velocity as a function of time for simulations S15s7b\_nt, 
S15s7b\_hyb, S25s7b\_nt, and S25s7b\_hyb. The dashed lines refer to the simulations with Newtonian
hydrodynamics (S15s7b\_nt and S25s7b\_nt), and the solid lines refer to the simulations with GR
hydrodynamics (S15s7b\_hyb and S25s7b\_hyb). All simulations utilize Newtonian transport. 
\label{fig3}} 

\figcaption[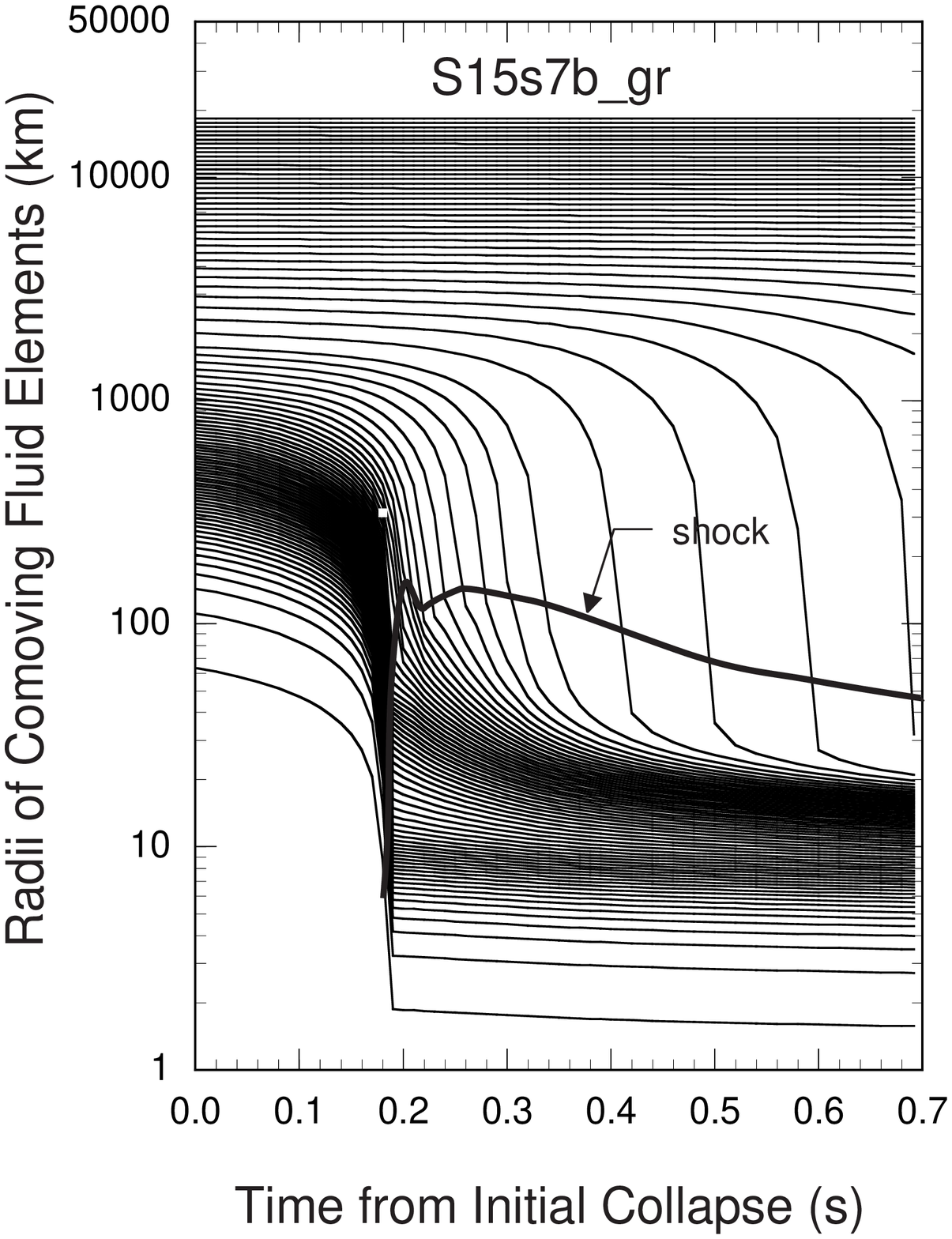] {Radii as a function of time of spherical shells at selected
rest masses for the fully GR simulation, S15s7b\_gr, for the first 0.69 s.  \label{fig4}} 

\figcaption[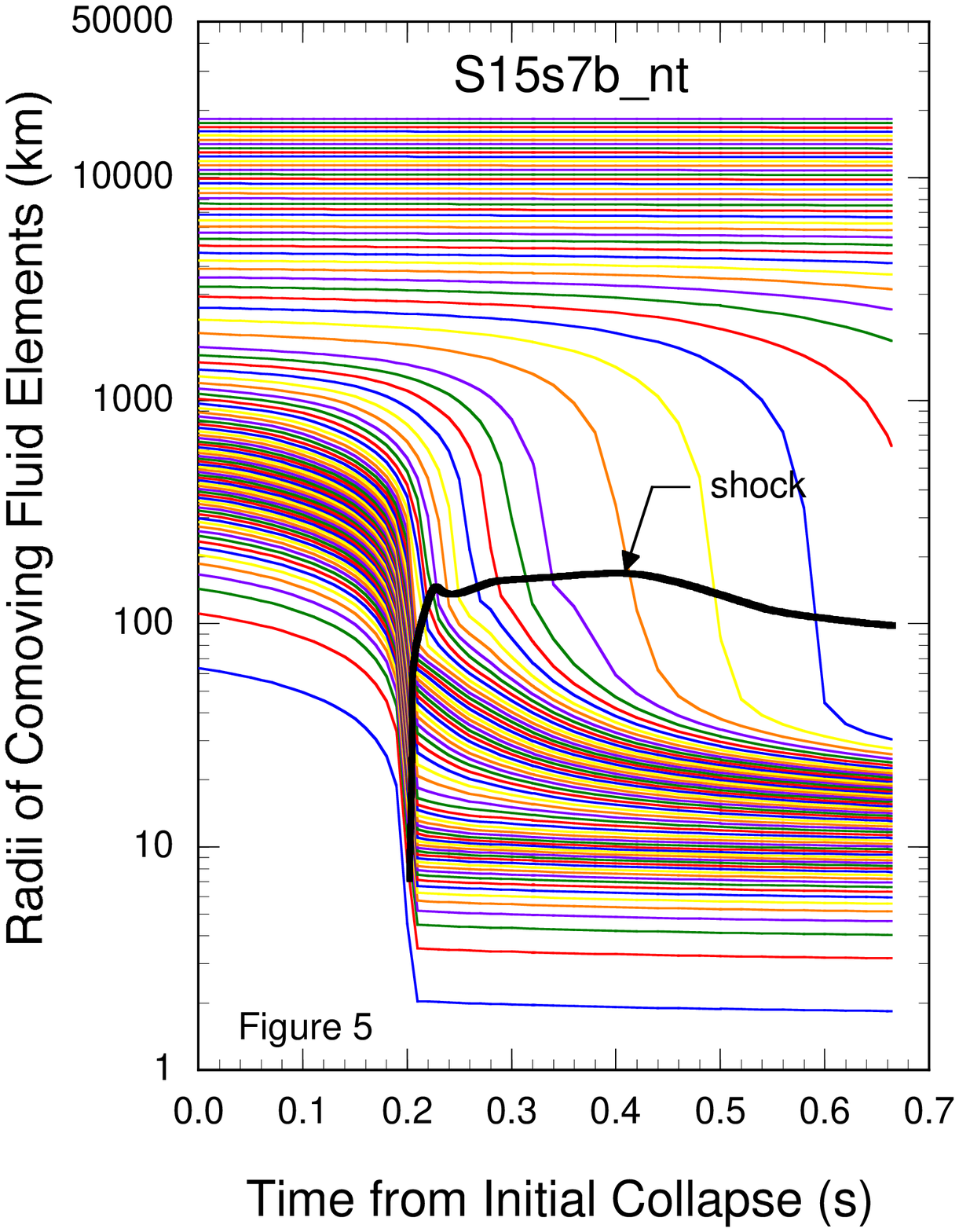] {Radii as a function of time of spherical shells at selected
rest masses for the Newtonian simulation, S15s7b\_nt, for the first 0.65 s.  \label{fig5}}

\figcaption[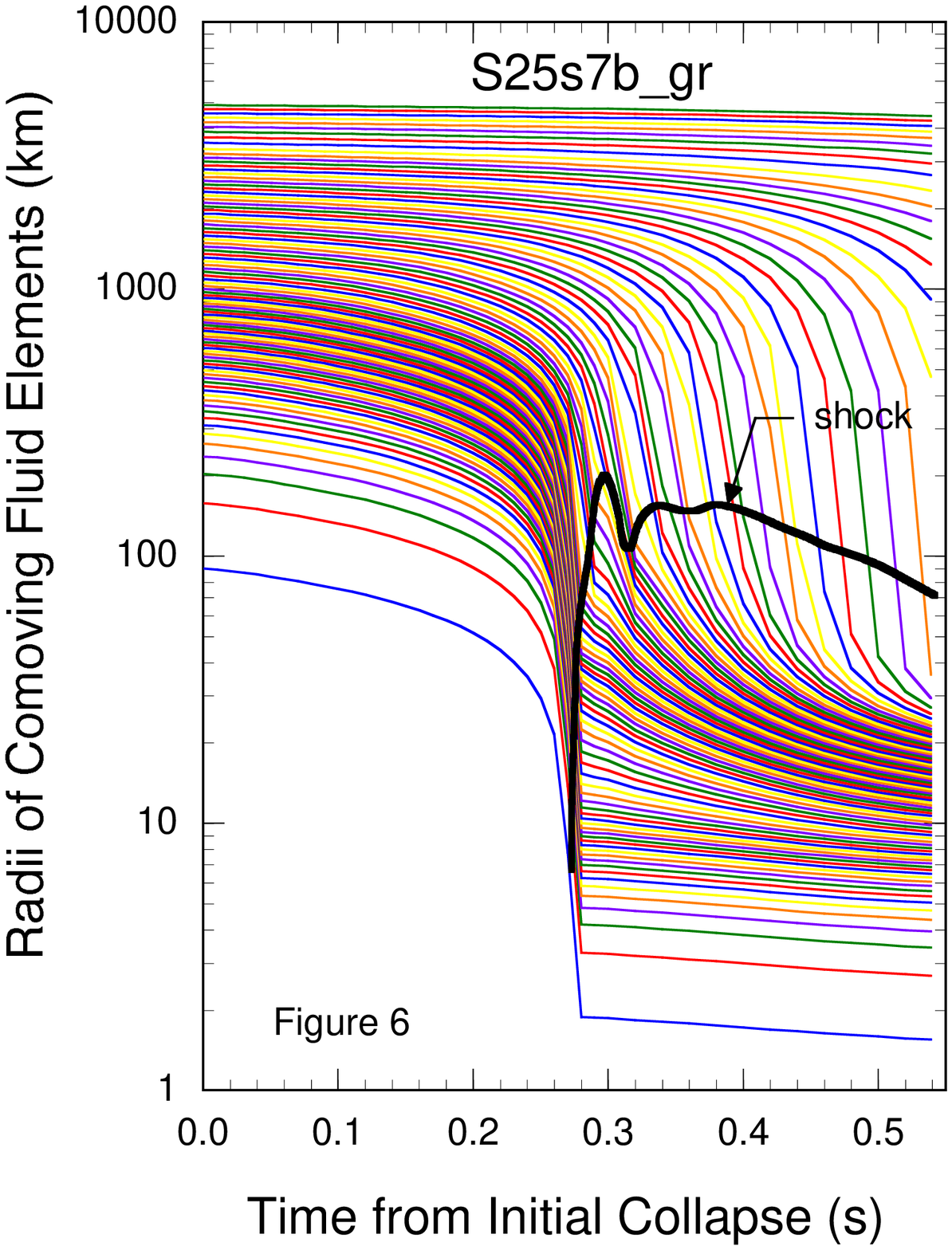] {Radii as a function of time of spherical shells at selected
rest masses for the fully GR simulation, S25s7b\_gr, for the first 0.55 s.  \label{fig6}} 

\figcaption[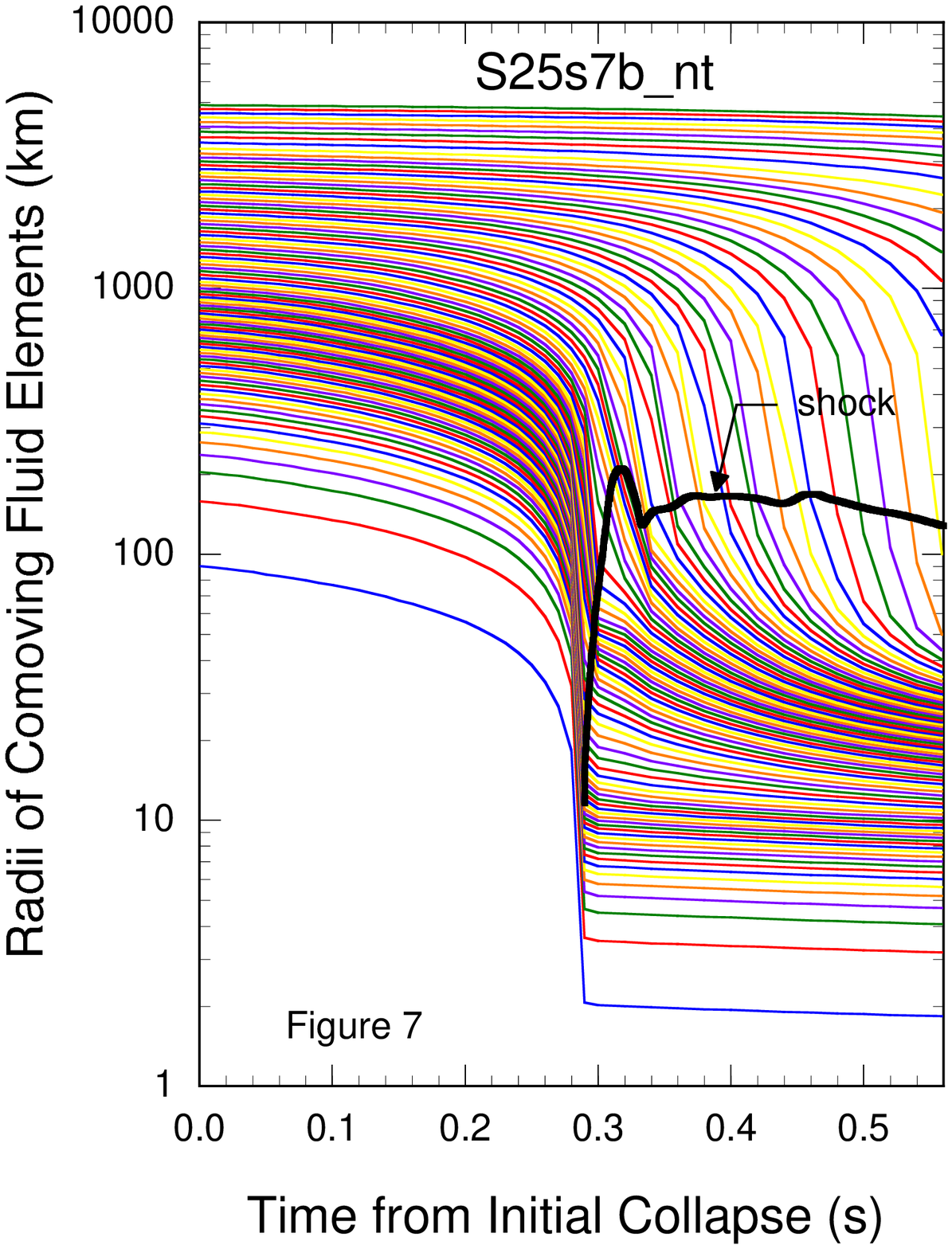] {Radii as a function of time of spherical shells at selected
rest masses for the fully Newtonian simulation, S25s7b\_nt, for the first 0.56 s. 
\label{fig7}} 

\clearpage

\figcaption[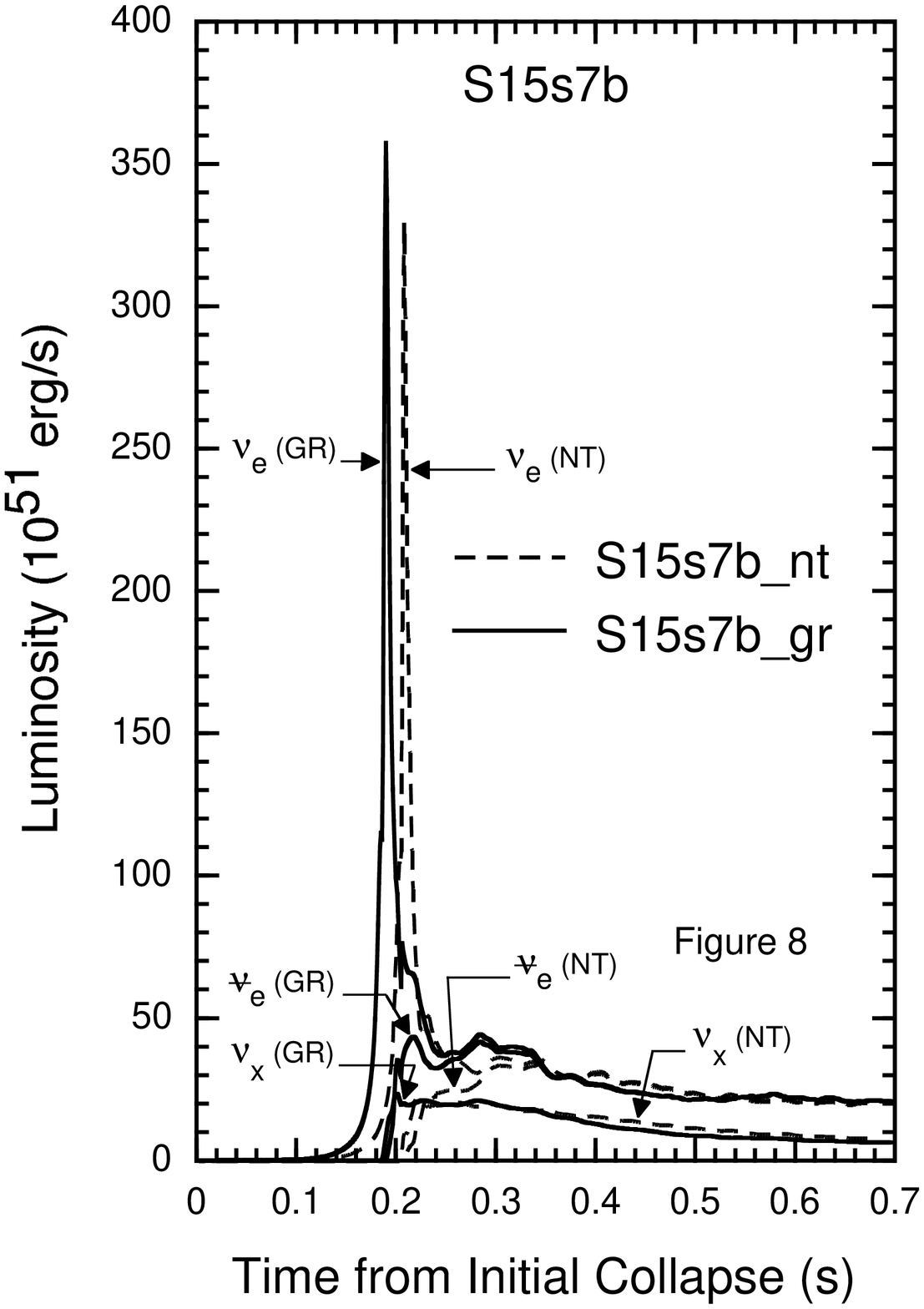] {Neutrino luminosities for each neutrino flavor as a function of time for
simulations S15s7b\_gr (fully GR) and S15s7b\_nt (Newtonian) for the first 0.7 s. The
$\nu_{\mu}$'s, $\nu_{\tau}$'s and their antiparticles are treated identically in the code, and the
luminosity of each is shown by the curves denoted by $\nu_{\rm x}$. Solid curves refer to
simulation S15s7b\_gr and are denoted by ``(GR)''; dashed curves refer to simulation S15s7b\_gr and
are denoted by ``(NT)''.
\label{fig8}} 

\figcaption[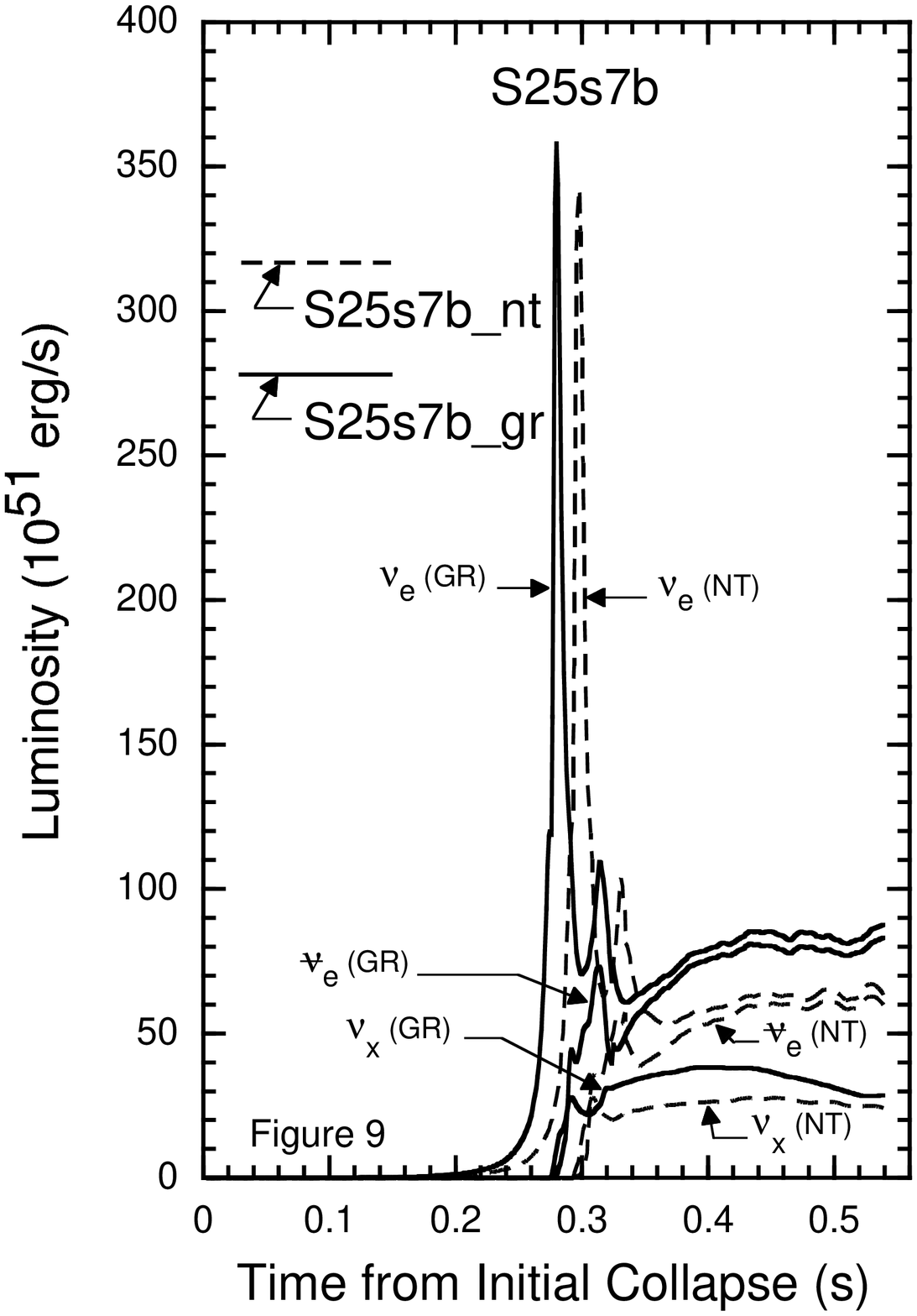] {The same quantities shown in Figure 8 but for simulations S25s7b\_nt and
S25s7b\_gr for the first 0.55 s. \label{fig9}} 

\figcaption[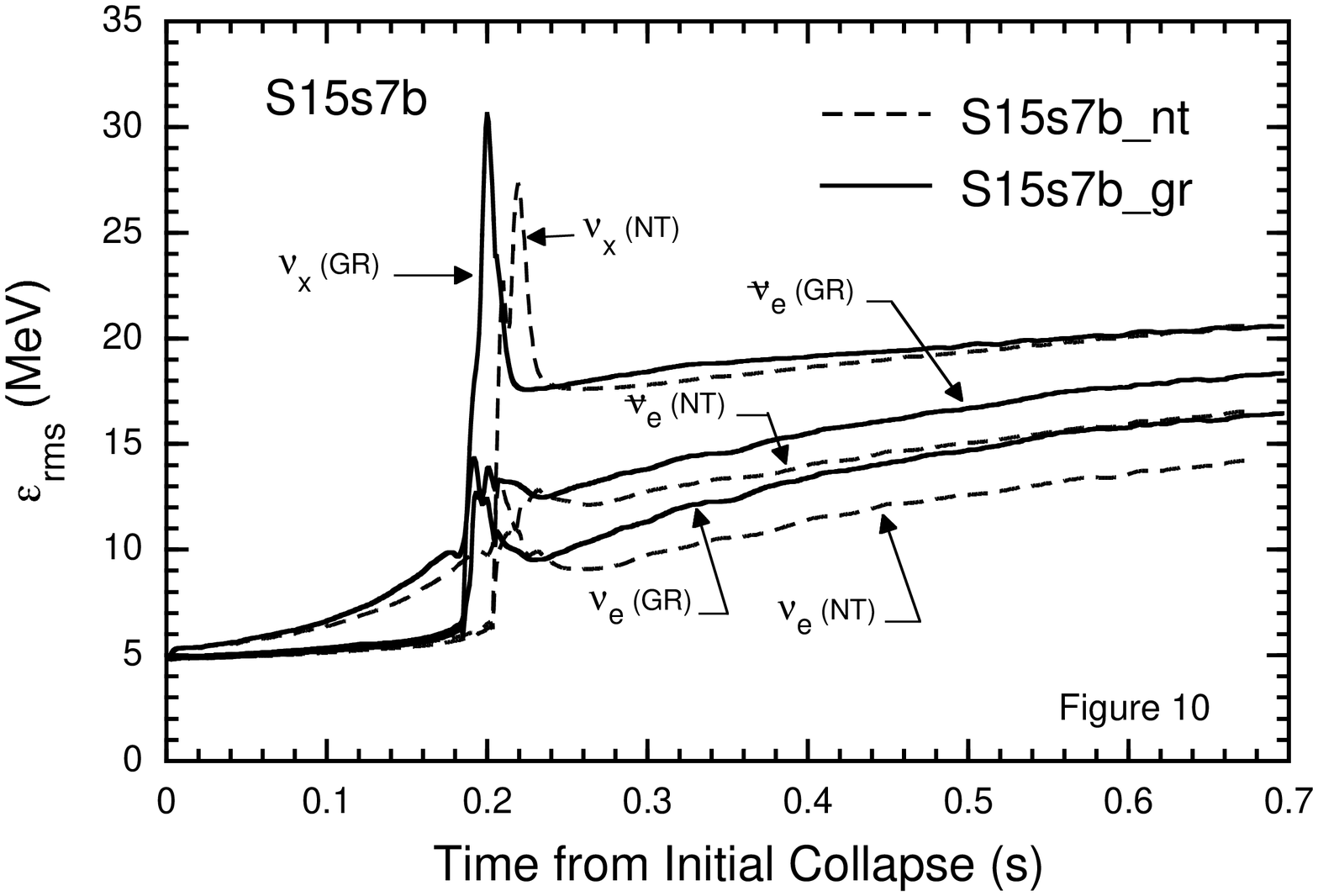] {Neutrino rms energies for each neutrino flavor as a function of time for
simulations S15s7b\_gr (fully GR) and S15s7b\_nt (Newtonian) for the first 0.7 s. The
$\nu_{\mu}$'s, $\nu_{\tau}$'s and their antiparticles are treated identically in the code, and the
rms energies of each is shown by the curves denoted by $\nu_{\rm x}$. Solid curves refer to
simulation S15s7b\_gr and are denoted by ``(GR)''; dashed curves refer to simulation S15s7b\_gr and
are denoted by ``(NT)''. \label{fig10}} 

\figcaption[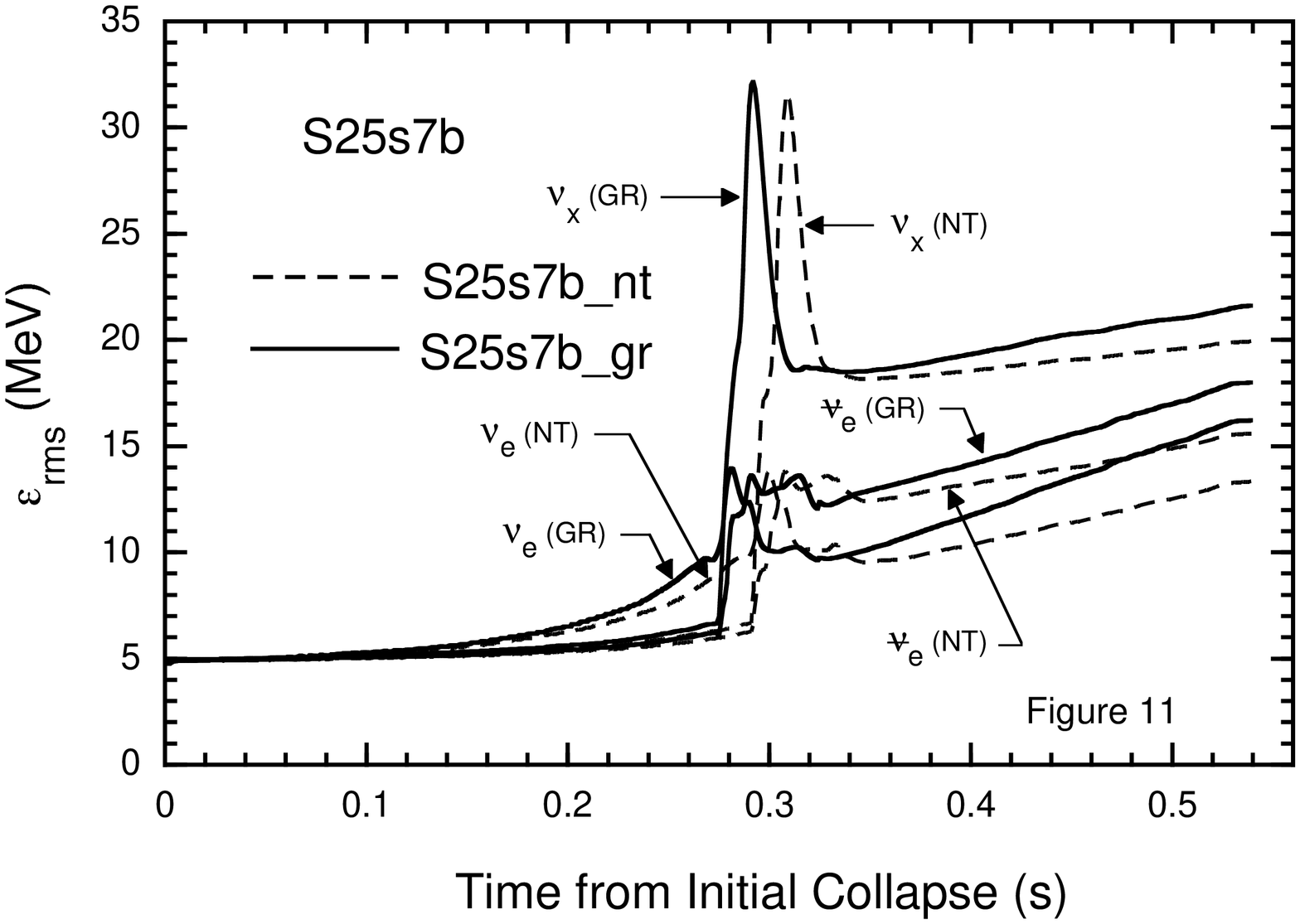] {The same quantities shown in Figure 10, but for simulations S25s7b\_nt and
S25s7b\_gr for the first 0.55 s. \label{fig11}} 

\figcaption[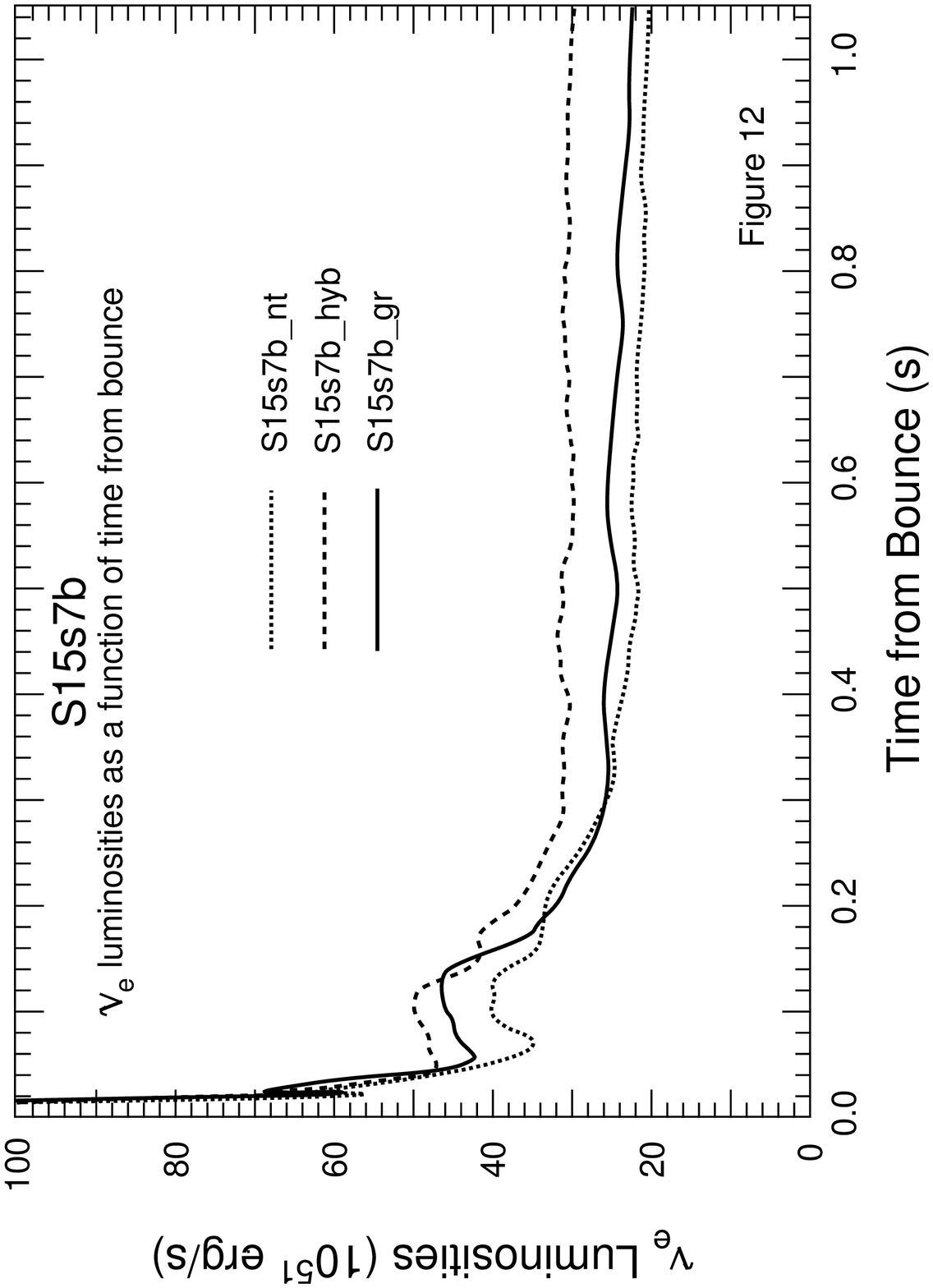] {Comparison of the $\nu_{\rm e}$ luminosities as a function of time from
bounce for simulation S15s7b\_nt (dotted line), simulation S15s7b\_hyb (dashed line), and
simulation S15s7b\_gr (solid line). \label{fig12}} 

\figcaption[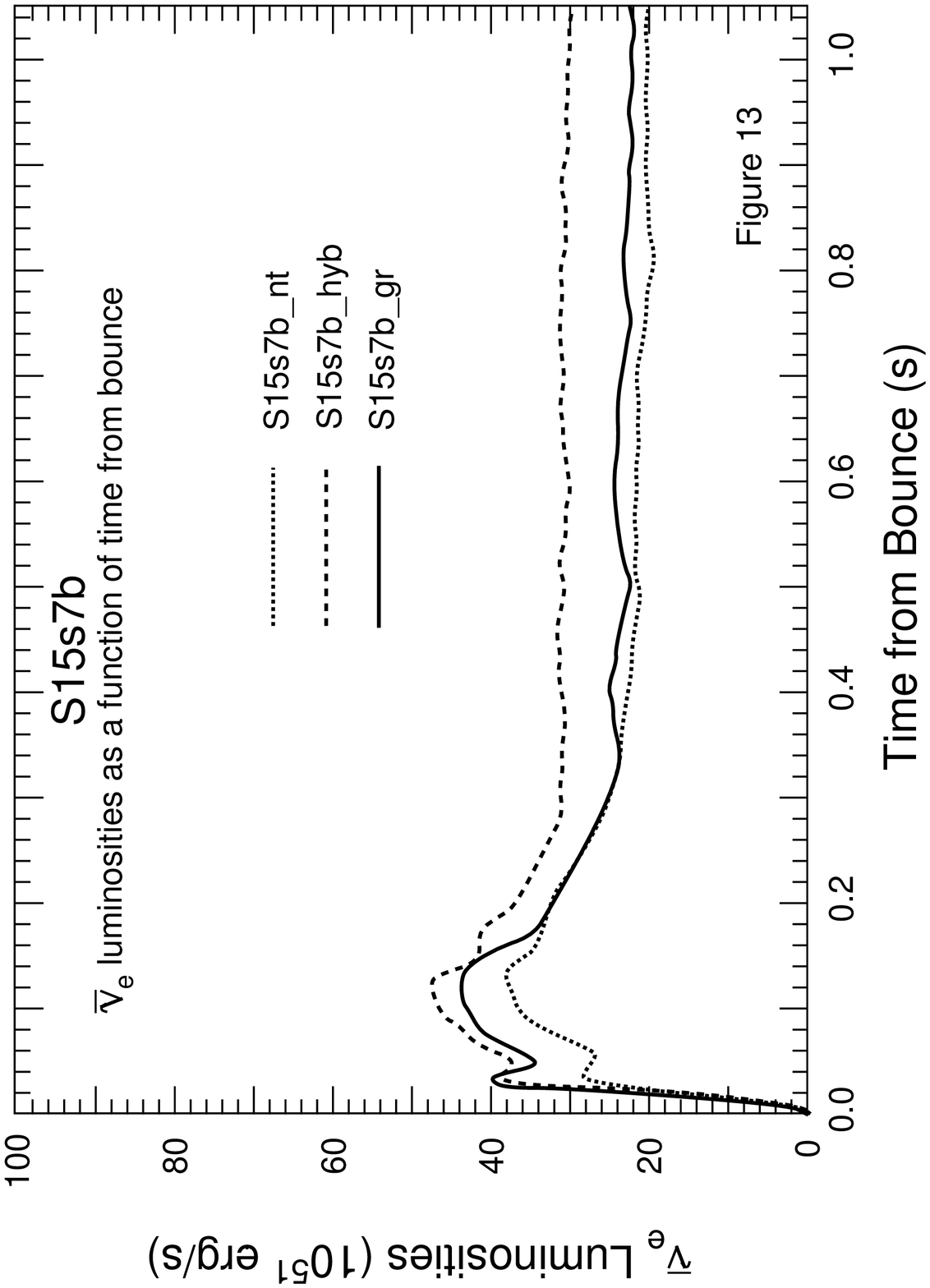] {Comparison of the $\bar{\nu}_{\rm e}$ luminosities as a function of time
from bounce for simulation S15s7b\_nt (dotted line), simulation S15s7b\_hyb (dashed line), and
simulation S15s7b\_gr (solid line). \label{fig13}} 

\figcaption[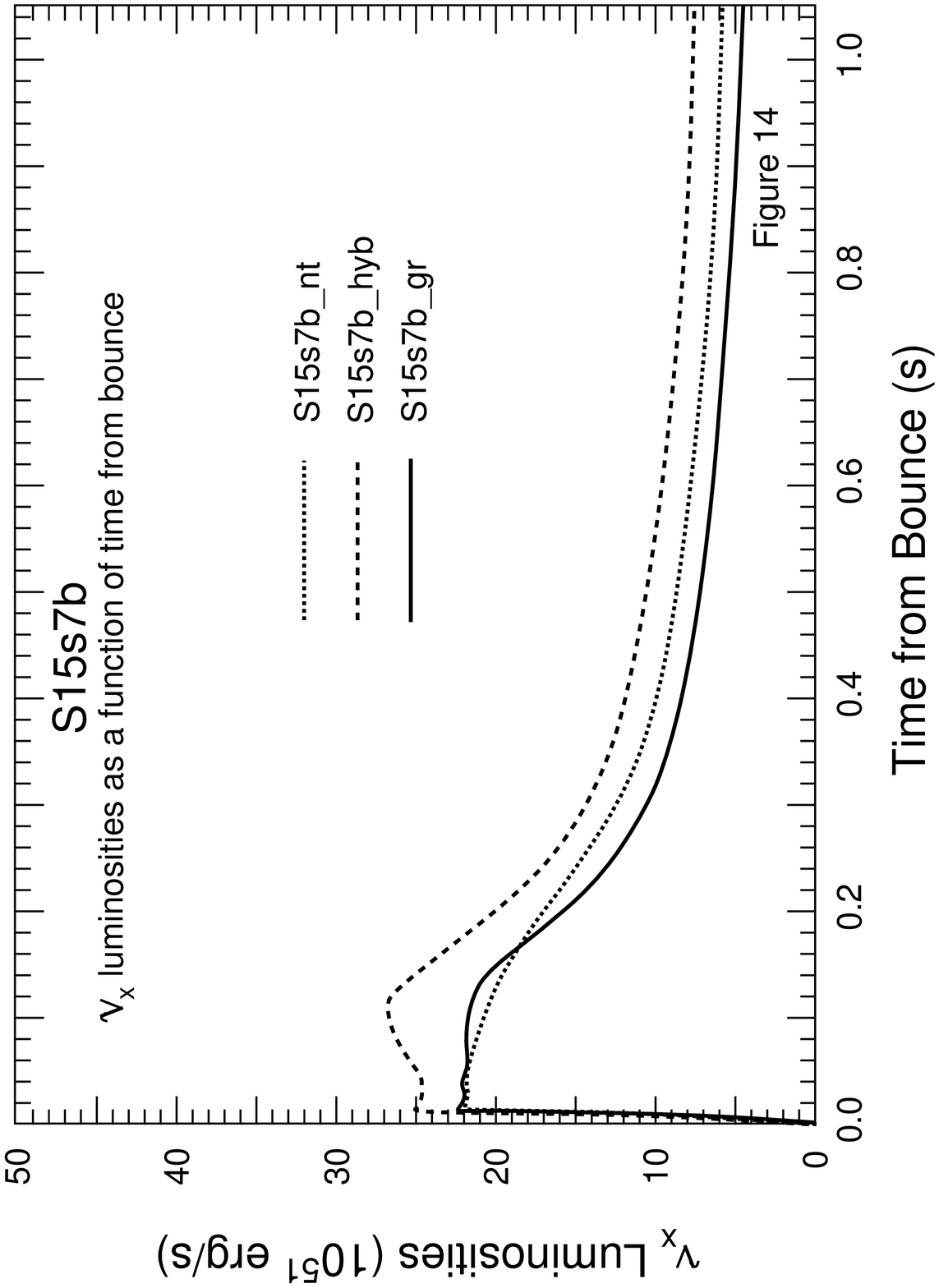] {Comparison of the $\bar{\nu}_{\rm x}$ luminosities as a function of time
from bounce for for simulation S15s7b\_nt (dotted line), simulation S15s7b\_hyb (dashed line), and
simulation S15s7b\_gr (solid line). Here $\nu_{\rm x}$ refers to either the $\nu_{\mu}$'s,
$\nu_{\tau}$'s, or their respective antiparticles. \label{fig14}} 

\figcaption[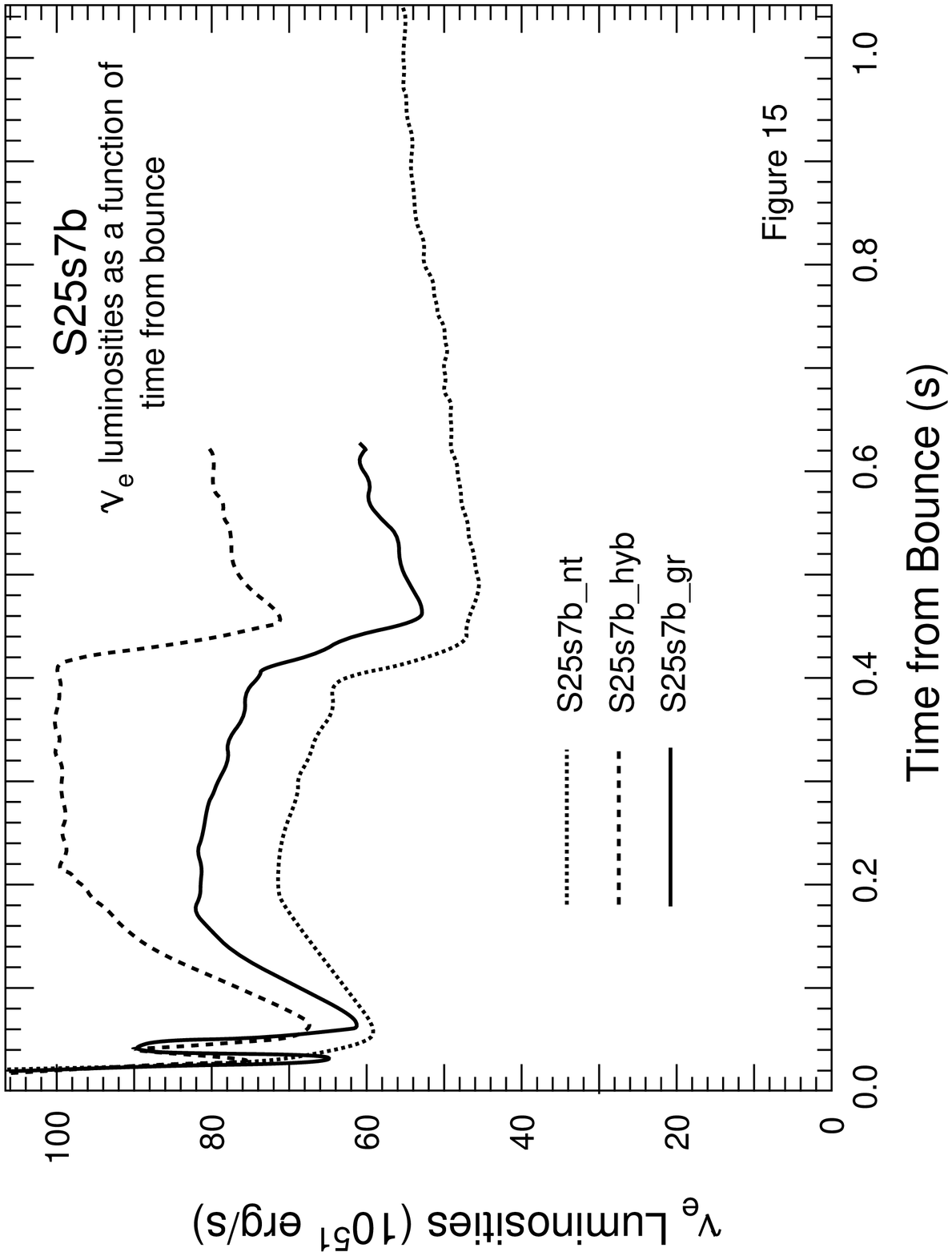] {Comparison of the $\nu_{\rm e}$ luminosities as a function of time from
bounce for simulation S25s7b\_nt (dotted line), simulation S25s7b\_hyb (dashed line), and
simulation S25s7b\_gr (solid line). \label{fig15}} 

\figcaption[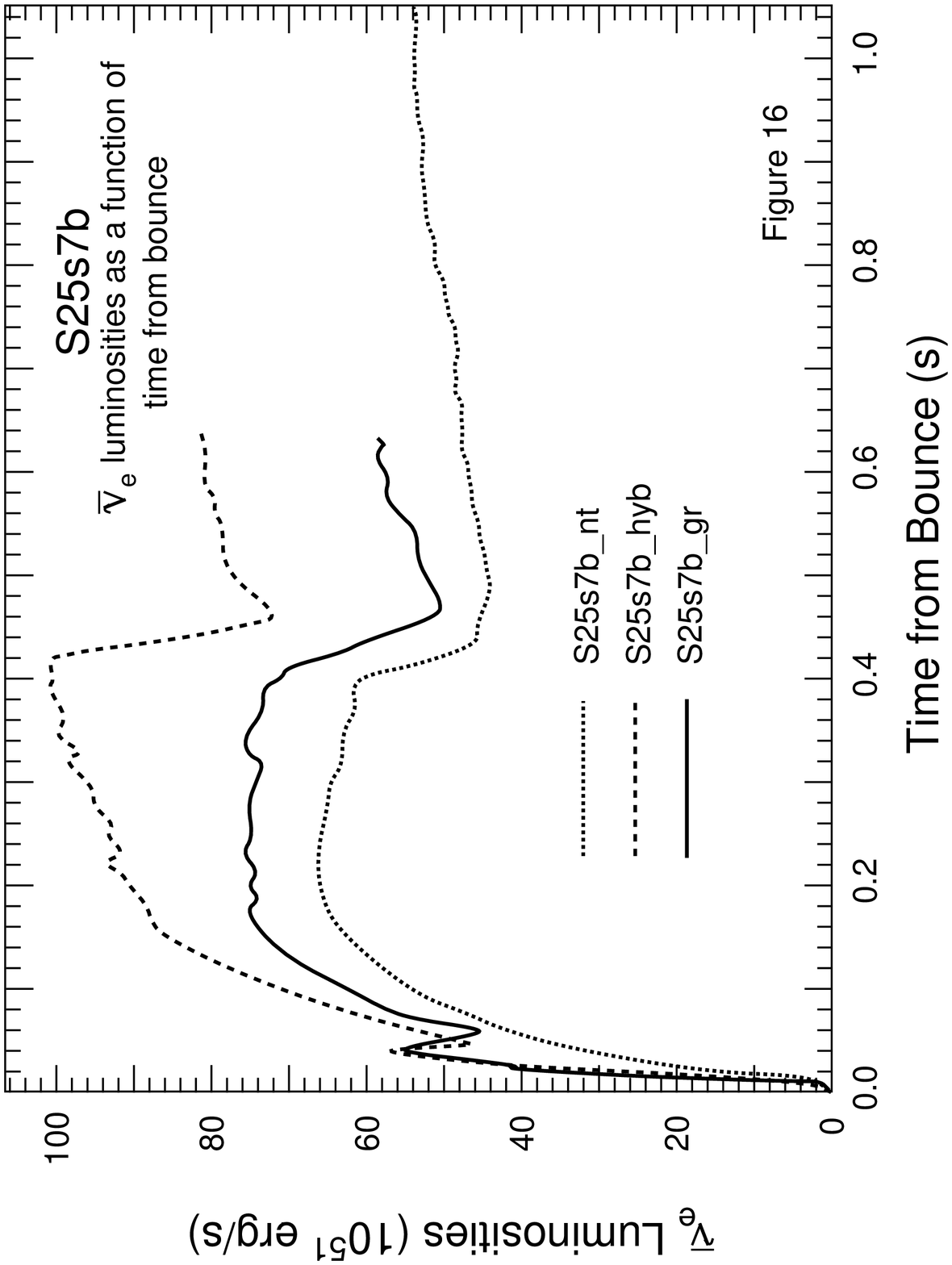] {Comparison of the $\bar{\nu}_{\rm e}$ luminosities as a function of time
from bounce for simulation S25s7b\_nt (dotted line), simulation S25s7b\_hyb (dashed line), and
simulation S25s7b\_gr (solid line). \label{fig16}}

\figcaption[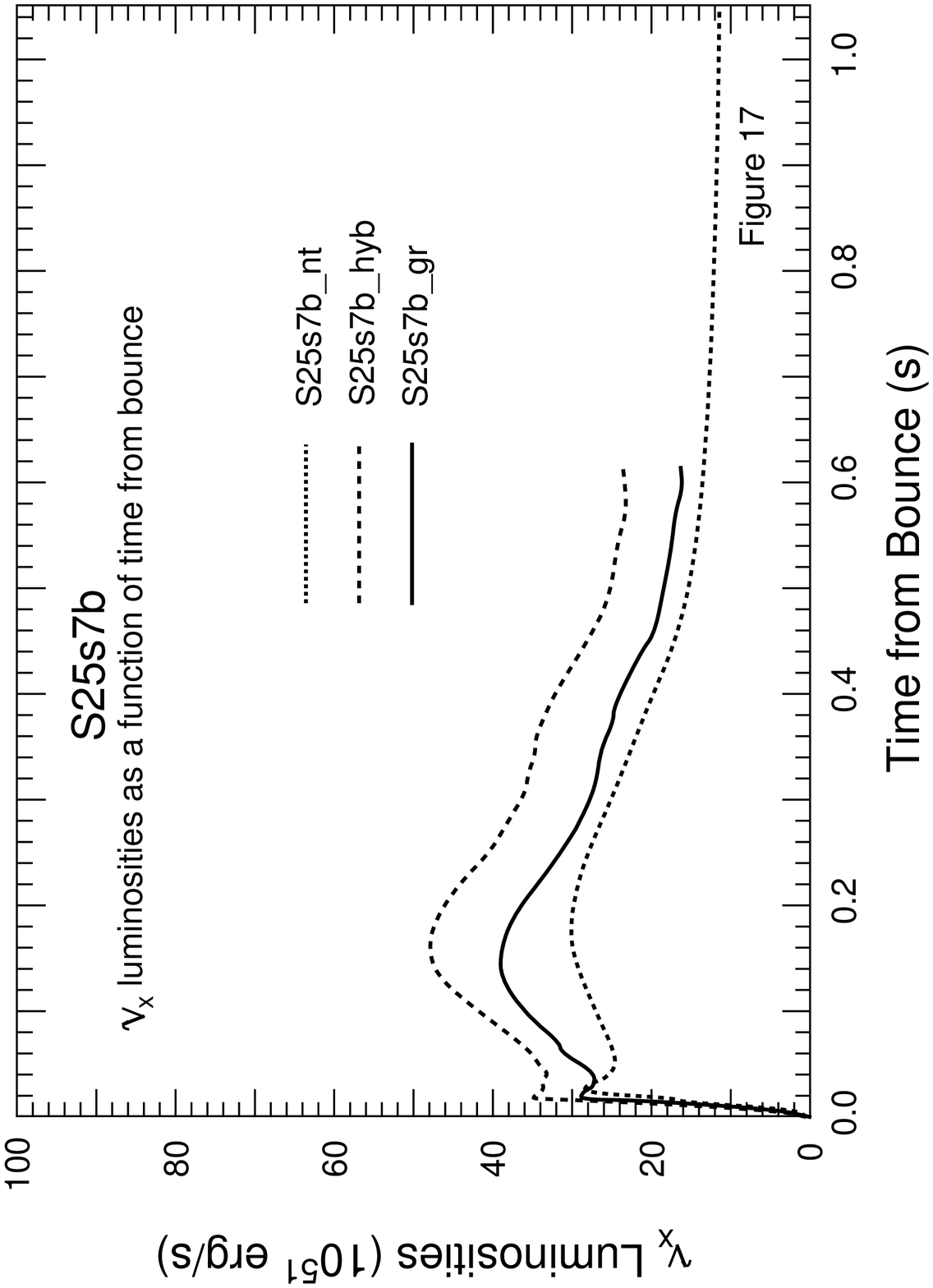] {Comparison of the $\bar{\nu}_{\rm x}$ luminosities as a function of time
from bounce for simulation S25s7b\_nt (dotted line), simulation S25s7b\_hyb (dashed line), and
simulation S25s7b\_gr (solid line). Here $\nu_{\rm x}$ refers to either the $\nu_{\mu}$'s,
$\nu_{\tau}$'s, or their respective antiparticles. \label{fig17}}

\figcaption[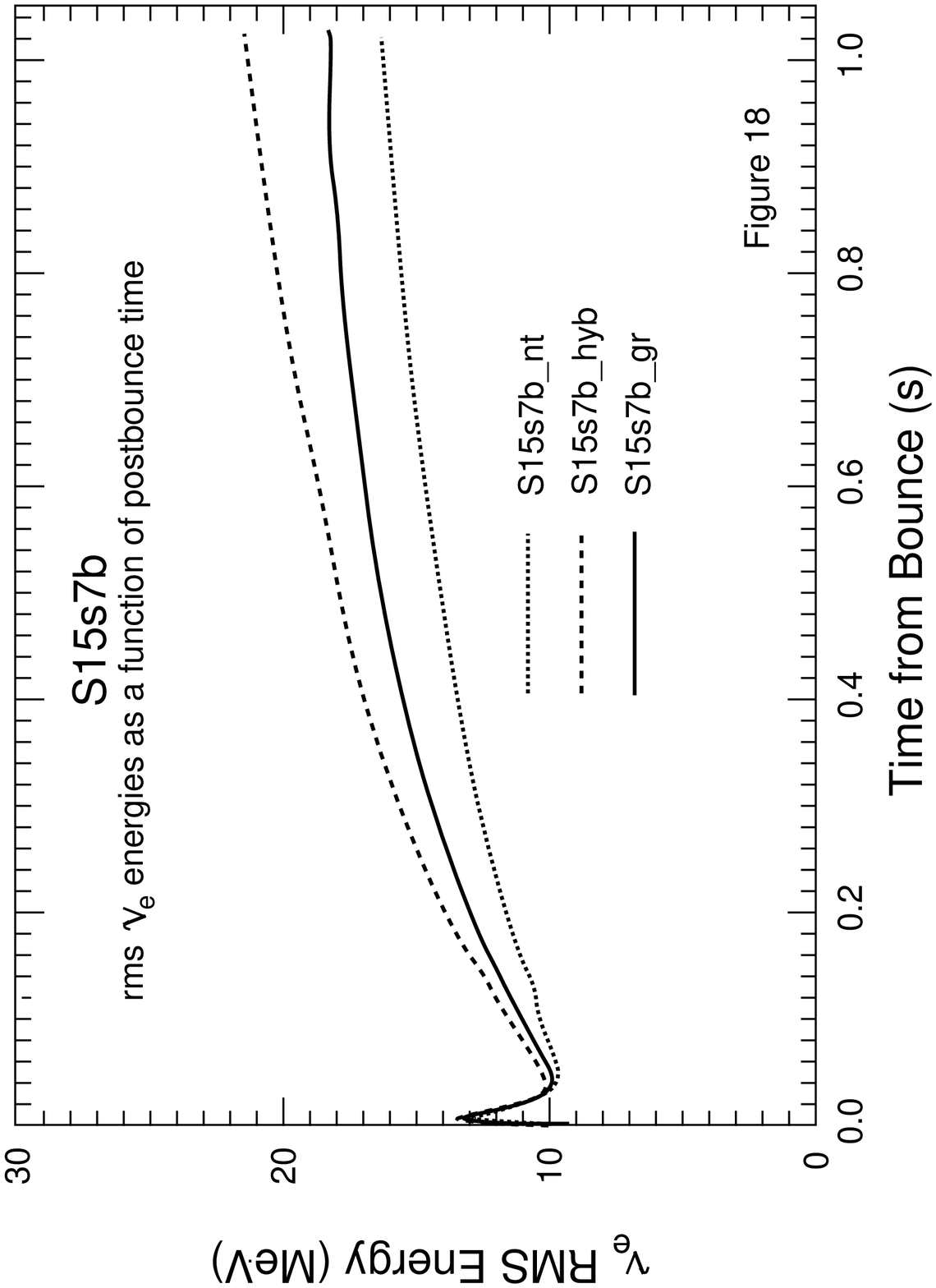] {Comparison of the $\nu_{\rm e}$ rms energies as a function of time from
bounce for simulation S15s7b\_nt (dotted line), simulation S15s7b\_hyb (dashed line), and
simulation S15s7b\_gr (solid line). \label{fig18}}

\figcaption[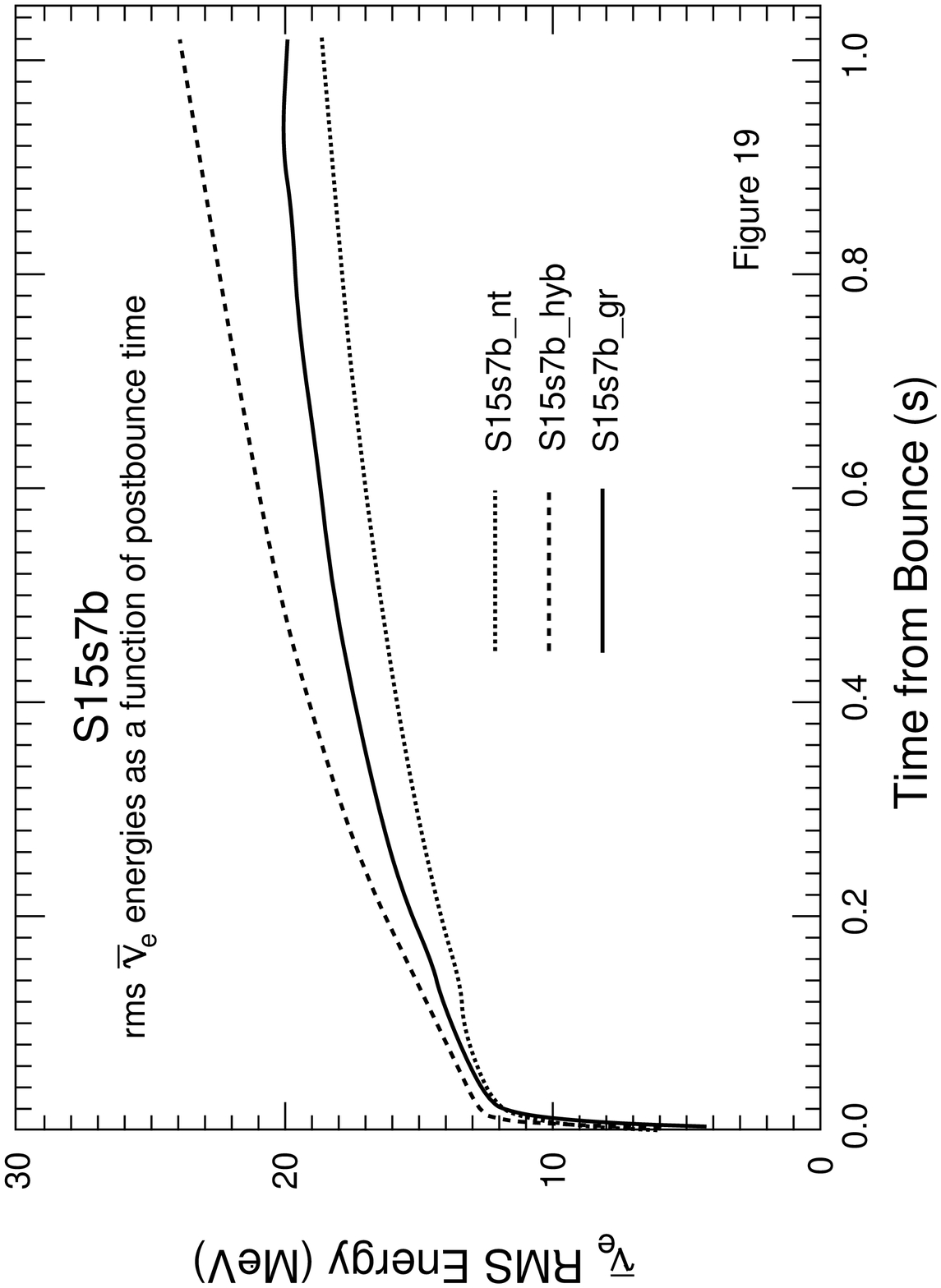] {Comparison of the $\bar{\nu}_{\rm e}$ rms energies as a function of time
from bounce for simulation S15s7b\_nt (dotted line), simulation S15s7b\_hyb (dashed line), and
simulation S15s7b\_gr (solid line). \label{fig19}}

\figcaption[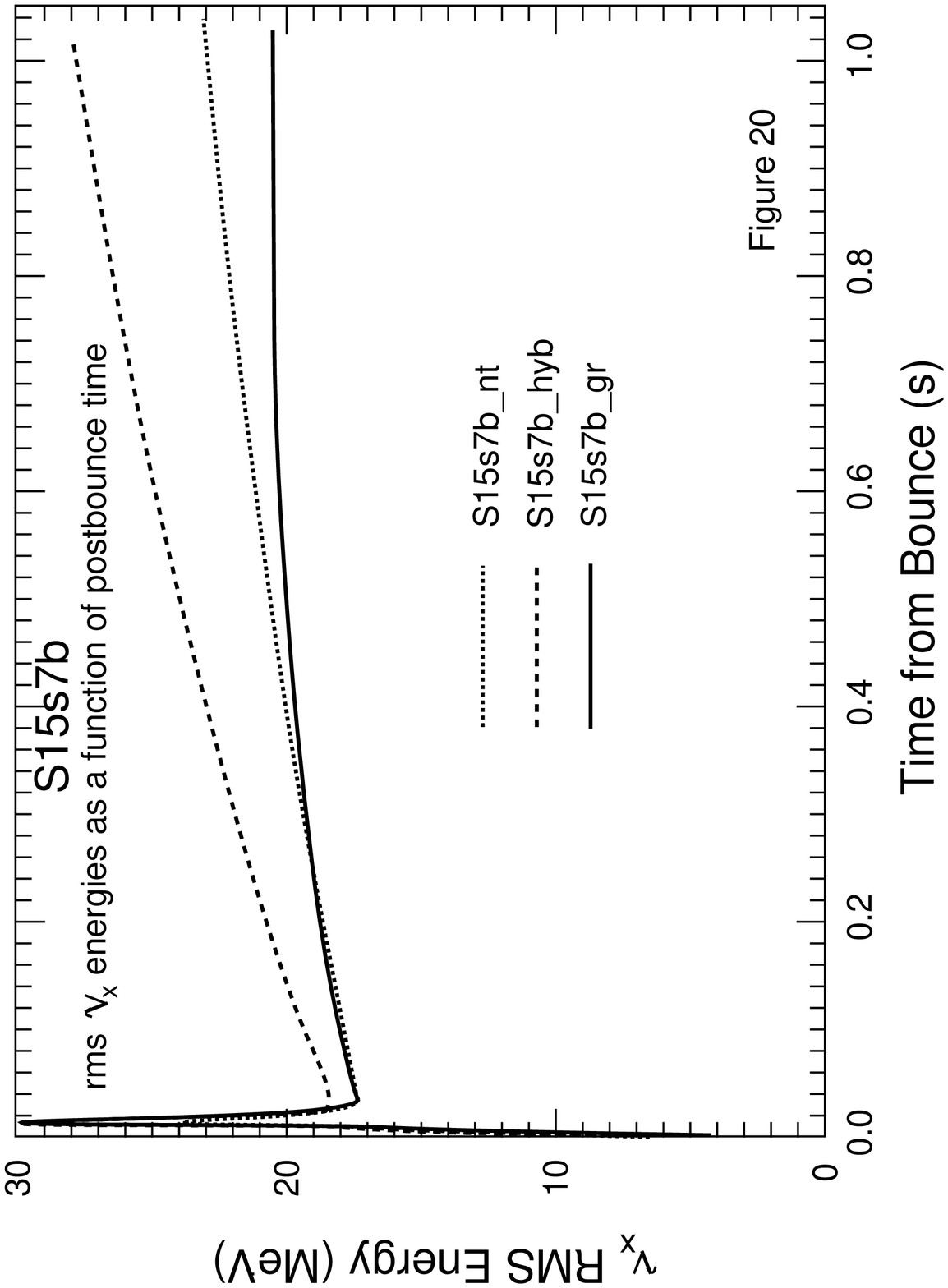] {Comparison of the $\bar{\nu}_{\rm x}$ rms energies as a function of time
from bounce for simulation S15s7b\_nt (dotted line), simulation S15s7b\_hyb (dashed line), and
simulation S15s7b\_gr (solid line). Here $\nu_{\rm x}$ refers to either the 
$\nu_{\mu}$'s, $\nu_{\tau}$'s, or their respective antiparticles. \label{fig20}}

\figcaption[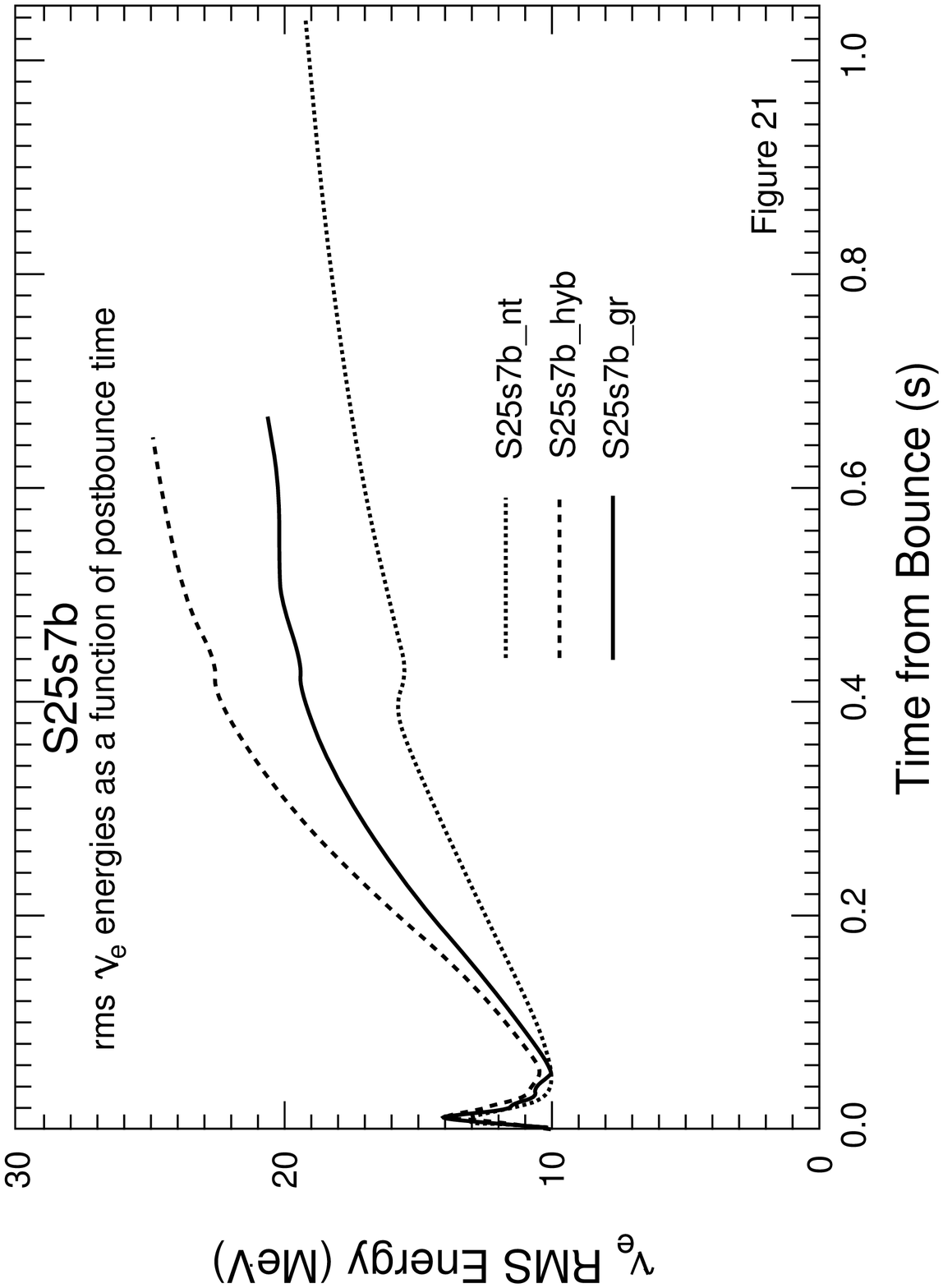] {Comparison of the $\nu_{\rm e}$ rms energies as a function of time from
bounce for simulation S25s7b\_nt (dotted line), simulation S25s7b\_hyb (dashed line), and
simulation S25s7b\_gr (solid line). \label{fig21}}

\figcaption[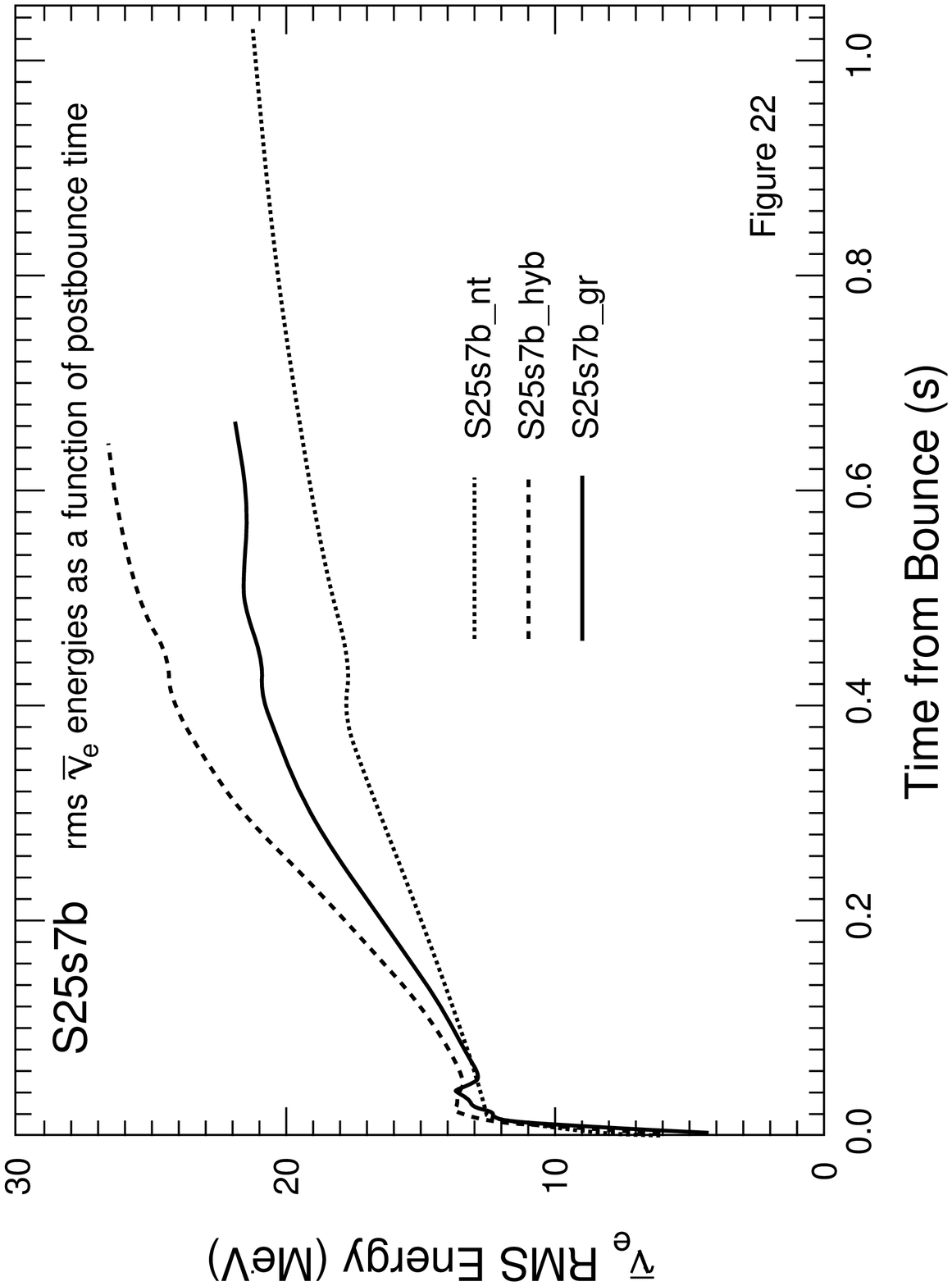] {Comparison of the $\bar{\nu}_{\rm e}$ rms energies as a function of
time from bounce for simulation S25s7b\_nt (dotted line), simulation S25s7b\_hyb (dashed line), and
simulation S25s7b\_gr (solid line). \label{fig22}}

\figcaption[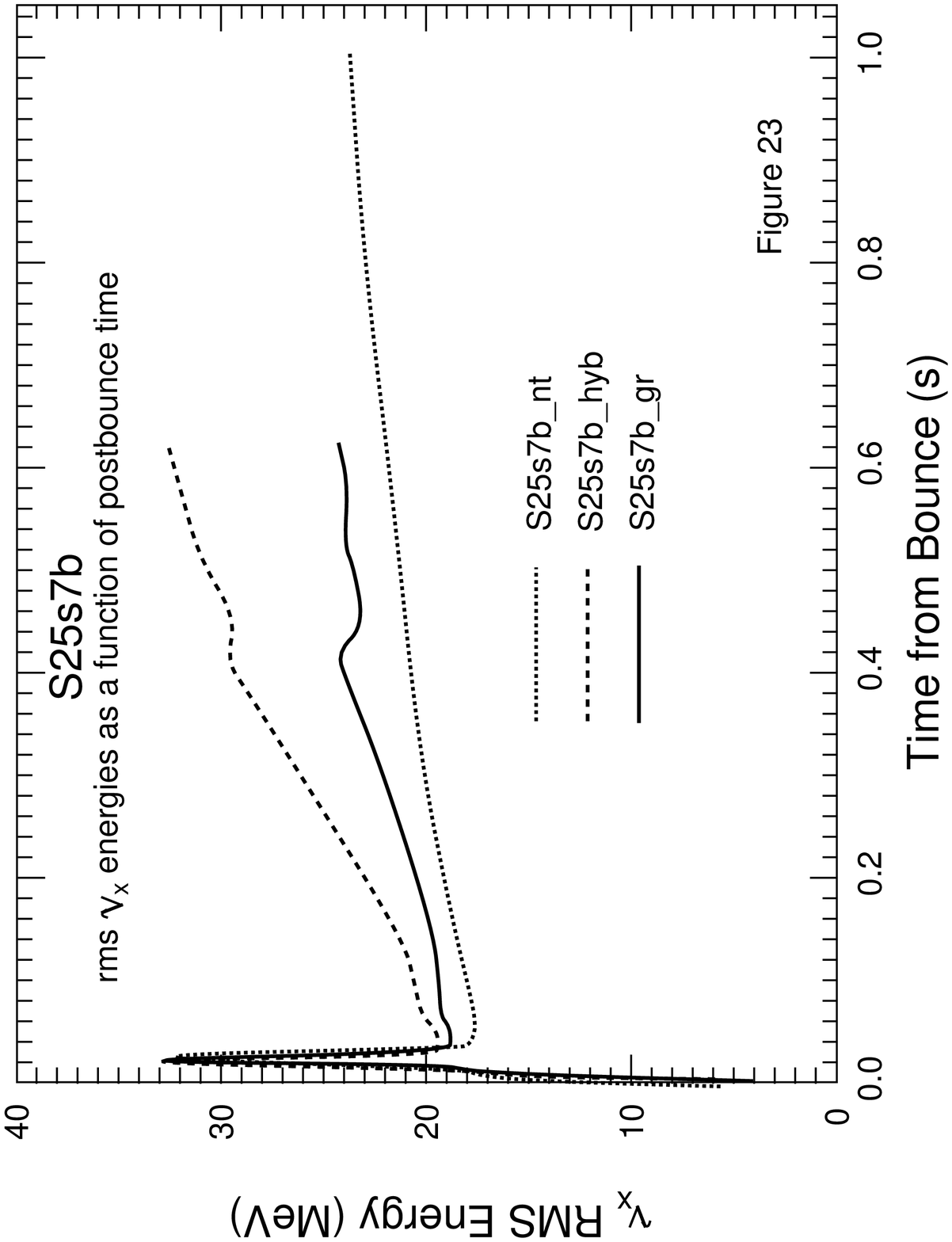] {Comparison of the $\bar{\nu}_{\rm x}$ rms energies as a function of time
from bounce for simulation S25s7b\_nt (dotted line), simulation S25s7b\_hyb (dashed line), and
simulation S25s7b\_gr (solid line). Here $\nu_{\rm x}$ refers to either the $\nu_{\mu}$'s,
$\nu_{\tau}$'s, or their respective antiparticles. \label{fig23}} 

\newpage
\newpage

\begin{figure}[t]
\setlength{\unitlength}{1cm}
\hspace*{0.1cm}
{\includegraphics[scale = 0.8]{Fig1.eps}}
%\caption{\label
%{fig:Fig1}Fig1}
%\figrule
\end{figure}

\newpage

\begin{figure}[t]
\setlength{\unitlength}{1cm}
\hspace*{0.1cm}
{\includegraphics[scale = 0.8]{Fig2.eps}}
%\caption{\label
%{fig:Fig2}Fig2}
%\figrule
\end{figure}

\newpage

\begin{figure}[t]
\setlength{\unitlength}{1cm}
\hspace*{0.1cm}
{\includegraphics[scale = 0.8]{Fig3.eps}}
%\caption{\label
%{fig:Fig3}Fig3}
%\figrule
\end{figure}

\newpage

\begin{figure}[t]
\setlength{\unitlength}{1cm}
\vspace*{-0.8cm}
\hspace*{-1.0cm}
{\includegraphics[scale = 0.8]{Fig4.eps}}
%\caption{\label
%{fig:Fig4}Fig4}
%\figrule
\end{figure}

\newpage

\begin{figure}[t]
\setlength{\unitlength}{1cm}
\vspace*{-0.8cm}
\hspace*{-1.0cm}
{\includegraphics[scale = 0.8]{Fig5.eps}}
%\caption{\label
%{fig:Fig5}Fig5}
%\figrule
\end{figure}

\newpage

\begin{figure}[t]
\setlength{\unitlength}{1cm}
\vspace*{-0.8cm}
\hspace*{-1.0cm}
{\includegraphics[scale = 0.8]{Fig6.eps}}
%\caption{\label
%{fig:Fig6}Fig6}
%\figrule
\end{figure}

\newpage

\begin{figure}[t]
\setlength{\unitlength}{1cm}
\vspace*{-0.8cm}
\hspace*{-1.0cm}
{\includegraphics[scale = 0.8]{Fig7.eps}}
%\caption{\label
%{fig:Fig7}Fig7}
%\figrule
\end{figure}

\newpage

\begin{figure}[t]
\setlength{\unitlength}{1cm}
\vspace*{-0.8cm}
\hspace*{-1.0cm}
{\includegraphics[scale = 0.8]{Fig8.eps}}
%\caption{\label
%{fig:Fig8}Fig8}
%\figrule
\end{figure}

\newpage

\begin{figure}[t]
\setlength{\unitlength}{1cm}
\vspace*{-0.8cm}
\hspace*{-1.0cm}
{\includegraphics[scale = 0.8]{Fig9.eps}}
%\caption{\label
%{fig:Fig9}Fig9}
%\figrule
\end{figure}

\newpage

\begin{figure}[t]
\setlength{\unitlength}{1cm}
\vspace*{-0.8cm}
\hspace*{-2.0cm}
{\includegraphics[scale = 0.7]{Fig10.eps}}
%\caption{\label
%{fig:Fig10}Fig10}
%\figrule
\end{figure}

\newpage

\begin{figure}[t]
\setlength{\unitlength}{1cm}
\vspace*{-0.8cm}
\hspace*{-2.0cm}
{\includegraphics[scale = 0.7]{Fig11.eps}}
%\caption{\label
%{fig:Fig11}Fig11}
%\figrule
\end{figure}

\newpage

\begin{figure}[t]
\setlength{\unitlength}{1cm}
\hspace*{0.1cm}
{\includegraphics[scale = 0.8]{Fig12.eps}}
%\caption{\label
%{fig:Fig12}Fig12}
%\figrule
\end{figure}

\newpage

\begin{figure}[t]
\setlength{\unitlength}{1cm}
\hspace*{0.1cm}
{\includegraphics[scale = 0.8]{Fig13.eps}}
%\caption{\label
%{fig:Fig13}Fig13}
%\figrule
\end{figure}

\newpage

\begin{figure}[t]
\setlength{\unitlength}{1cm}
\hspace*{0.1cm}
{\includegraphics[scale = 0.8]{Fig14.eps}}
%\caption{\label
%{fig:Fig14}Fig14}
%\figrule
\end{figure}

\newpage

\begin{figure}[t]
\setlength{\unitlength}{1cm}
\hspace*{0.1cm}
{\includegraphics[scale = 0.8]{Fig15.eps}}
%\caption{\label
%{fig:Fig15}Fig15}
%\figrule
\end{figure}

\newpage

\begin{figure}[t]
\setlength{\unitlength}{1cm}
\hspace*{0.1cm}
{\includegraphics[scale = 0.8]{Fig16.eps}}
%\caption{\label
%{fig:Fig16}Fig16}
%\figrule
\end{figure}

\newpage

\begin{figure}[t]
\setlength{\unitlength}{1cm}
\hspace*{0.1cm}
{\includegraphics[scale = 0.8]{Fig17.eps}}
%\caption{\label
%{fig:Fig17}Fig17}
%\figrule
\end{figure}

\newpage

\begin{figure}[t]
\setlength{\unitlength}{1cm}
\hspace*{0.1cm}
{\includegraphics[scale = 0.8]{Fig18.eps}}
%\caption{\label
%{fig:Fig18}Fig18}
%\figrule
\end{figure}

\clearpage

\begin{figure}[h]
\setlength{\unitlength}{1cm}
\hspace*{0.1cm}
{\includegraphics[scale = 0.8]{Fig19.eps}}
%\caption{\label
%{fig:Fig19}Fig19}
%\figrule
\end{figure}

\newpage

\begin{figure}[t]
\setlength{\unitlength}{1cm}
\hspace*{0.1cm}
{\includegraphics[scale = 0.8]{Fig20.eps}}
%\caption{\label
%{fig:Fig20}Fig20}
%\figrule
\end{figure}

\newpage

\begin{figure}[t]
\setlength{\unitlength}{1cm}
\hspace*{0.1cm}
{\includegraphics[scale = 0.8]{Fig21.eps}}
%\caption{\label
%{fig:Fig21}Fig21}
%\figrule
\end{figure}

\newpage

\begin{figure}[t]
\setlength{\unitlength}{1cm}
\hspace*{0.1cm}
{\includegraphics[scale = 0.8]{Fig22.eps}}
%\caption{\label
%{fig:Fig22}Fig22}
%\figrule
\end{figure}

\newpage

\begin{figure}[t]
\setlength{\unitlength}{1cm}
\hspace*{0.1cm}
{\includegraphics[scale = 0.8]{Fig23.eps}}
%\caption{\label
%{fig:Fig23}Fig23}
%\figrule
\end{figure}

\end{document}